\documentclass{aa}

\usepackage{txfonts}
\usepackage{graphicx}

\begin{document}

\title{Effelsberg \ion{H}{i} observations of compact high-velocity clouds}

\author{T. Westmeier \and C. Br\"uns \and J. Kerp}
\offprints{T. Westmeier,\\ \email{twestmei@astro.uni-bonn.de}}
\institute{Radioastronomisches Institut der Universit\"at Bonn, Auf dem H\"ugel 71, 53121 Bonn, Germany}

\date{Received 7 May 2004 / Accepted 24 November 2004}

\abstract{We have mapped 11 compact high-velocity clouds (CHVCs) in the 21-cm line emission of neutral, atomic hydrogen, using 
  the Effelsberg 100-m radio telescope. The aim of our observations was to study the overall distribution of the warm neutral 
  medium of CHVCs with high sensitivity. The achieved baseline rms of $\sigma_{\mathrm{rms}} \approx 50 \; \mathrm{mK}$ at the 
  original $2.6 \; \mathrm{km \, s^{-1}}$ velocity resolution allows us to search for evidence of ram-pressure interaction with 
  the ambient medium. In addition, we have obtained spectra along an appropriate axis across each CHVC with longer integration 
  times and denser angular sampling. These deep slices with $\sigma_{\mathrm{rms}} \approx 25 \ldots 35 \; \mathrm{mK}$ allow 
  us to determine the column density profile in greater detail as well as the velocity and line width gradient across each cloud.

  The most outstanding result of our observations is the complexity of the \ion{H}{i} column density distribution and the line 
  profiles of the investigated CHVCs. We have found only one cloud with a spherically-symmetric appearance. Among 
  the remaining clouds we observe head-tail structures, bow-shock shapes, and objects with irregular shapes. These complex 
  morphologies in combination with the obtained physical parameters suggest that ram-pressure interactions with an ambient 
  medium may play a significant role in shaping some of the CHVCs from our sample. These results are consistent with a 
  circumgalactic distribution of CHVCs with typical distances of the order of 100~kpc. The pressure of the ambient medium might 
  also stabilise CHVCs in addition to their own gravitational potential.
  } 

\maketitle

\begin{table*}
\caption{\label{chvc_table}Physical parameters of the 11 observed CHVCs. $l$ and $b$ are the Galactic longitude and 
latitude of the column density maximum, $\alpha$ and $\delta$ the corresponding equatorial coordinates, $v_{\mathrm{LSR}}$ 
and $v_{\mathrm{GSR}}$ the column density weighted average radial velocities in LSR and GSR frames, $\Delta v$ the average line 
width (FWHM), $T_{\mathrm{B}}$ the observed peak brightness temperature, $N_{\ion{H}{i}}$ the \ion{H}{i} peak column density, and 
$e$ the ellipticity of the cloud. The last two rows give the mean value $\langle x \rangle$ and the standard deviation $\sigma_x$ 
for each parameter $x$.}
\begin{center}
\begin{tabular}{lrrrrrrrr}
\hline
\hline
Name & $\alpha$(2000) & $\delta$(2000) & $v_{\mathrm{LSR}}$ & $v_{\mathrm{GSR}}$ & $\Delta v$ & $T_{\mathrm{B}}$ & $N_{\ion{H}{i}}$ & $e$ \\
(CHVC $l \pm b$) & & & ($\mathrm{km \, s^{-1}}$) & ($\mathrm{km \, s^{-1}}$) & ($\mathrm{km \, s^{-1}}$) & (K) & ($10^{19} \; {\mathrm{cm}^{-2}}$) & \\
\hline
CHVC 016.8$-$25.2   & $19^h 59^m 05^s$ & $-24^{\circ} 42'$ &    $-228$ &   $-171$ &    $14$ &     $3.1$ &     $5.6$ &	  $0.38$ \\
CHVC 032.1$-$30.7   & $20^h 41^m 40^s$ & $-14^{\circ} 07'$ &    $-308$ &   $-207$ &    $30$ &     $1.3$ &     $6.0$ &	  $0.33$ \\
CHVC 039.0$-$33.2   & $21^h 00^m 58^s$ & $-09^{\circ} 51'$ &    $-262$ &   $-147$ &    $22$ &     $1.6$ &     $8.0$ &	  $0.42$ \\
CHVC 039.9$+$00.6   & $19^h 02^m 03^s$ & $+06^{\circ} 28'$ &    $-278$ &   $-137$ &    $32$ &     $0.7$ &     $4.5$ &	  $0.49$ \\
CHVC 050.4$-$68.4   & $23^h 24^m 14^s$ & $-19^{\circ} 00'$ &    $-195$ &   $-133$ &    $27$ &     $1.3$ &     $4.7$ &	  $0.52$ \\
CHVC 147.5$-$82.3   & $01^h 05^m 02^s$ & $-20^{\circ} 05'$ &    $-269$ &   $-254$ &    $22$ &     $2.2$ &     $8.0$ &	  $0.32$ \\
CHVC 157.1$+$02.9   & $04^h 49^m 01^s$ & $+49^{\circ} 20'$ &    $-184$ &   $ -98$ &    $22$ &     $0.9$ &     $3.6$ &	  $0.58$ \\
CHVC 172.1$-$59.6   & $02^h 21^m 57^s$ & $-05^{\circ} 40'$ &    $-235$ &   $-219$ &    $28$ &     $0.9$ &     $4.2$ &	  $0.68$ \\
CHVC 218.1$+$29.0   & $08^h 44^m 18^s$ & $+08^{\circ} 41'$ &    $+145$ &   $ +27$ &    $ 6$ &     $2.8$ &     $3.2$ &	  $0.58$ \\
CHVC 220.5$-$88.2   & $00^h 59^m 28^s$ & $-27^{\circ} 21'$ &    $-258$ &   $-263$ &    $22$ &     $1.0$ &     $3.7$ &	  $0.58$ \\
CHVC 357.8$+$12.4   & $16^h 54^m 25^s$ & $-23^{\circ} 43'$ &    $-159$ &   $-167$ &    $27$ &     $1.5$ &     $6.4$ &	  $0.60$ \\
\hline
$\langle x \rangle$ &                  &                   &    $-203$ &   $-161$ &    $23$ &     $1.6$ &     $5.3$ &	  $0.50$ \\
$\sigma_x$          &                  &                   & $\pm 124$ & $\pm 81$ & $\pm 7$ & $\pm 0.8$ & $\pm 1.7$ & $\pm 0.12$ \\
\hline
\end{tabular}
\end{center}
\end{table*}

\section{Introduction}

High-velocity clouds (HVCs) were first detected by Muller et al. (\cite{muller}) in the 21-cm line emission of neutral, atomic 
hydrogen. HVCs are gas clouds which are characterised by high radial velocities incompatible with a participation in Galactic 
rotation. Over the past decades, there have been a large number of definitions concerning the velocities of HVCs. Following a 
suggested approach by Wakker (\cite{wakker}), their radial velocities have to be at least $50 \; \mathrm{km \, s^{-1}}$ higher than 
the maximum radial velocities allowed for Galactic gas in the same direction according to a simple model of Galactic rotation. 
Apart from several small clouds, HVCs appear as large, homogeneous complexes, some of them spanning tens of degrees across the sky 
(see Wakker \& van Woerden \cite{wakker2} for a detailed review).

Among the numerous hypotheses proposed for the origin and distribution of HVCs (Wakker \& van Woerden \cite{wakker2}), the Local 
Group hypothesis has become the most discussed during recent years. In 1999, Blitz et al. argued that the observed properties 
of HVCs were consistent with a distribution throughout the entire Local Group after excluding some of the large HVC complexes and 
the Magellanic Stream. Furthermore, they suggested that HVCs might be the so-called missing dark-matter satellites predicted by 
cosmological cold dark matter (CDM) models (Klypin et al. \cite{klypin}, Moore et al. \cite{moore}). These models predict many more 
dark-matter haloes for the Local Group than can be directly observed in the form of dwarf galaxies. Blitz et al. (\cite{blitz}) 
argued that these dark-matter haloes might not have formed stars on a grand scale and could, therefore, be identified with a 
population of HVCs spread across the entire Local Group.

Following the approach of Blitz et al. (\cite{blitz}), Braun \& Burton (\cite{braun}) defined a subclass of compact, isolated HVCs 
on the basis of the Leiden/Dwingeloo Survey of Galactic neutral hydrogen (LDS, Hartmann \& Burton \cite{hartmann}). They compiled 
a catalogue of 66 so-called compact high-velocity clouds (CHVCs), characterised by angular sizes of less than $2^{\circ}$ FWHM and 
spatial isolation and separation from neighbouring \ion{H}{i} emission. Braun \& Burton (\cite{braun}) found the statistical 
properties of these CHVCs to be consistent with a distribution throughout the entire Local Group with typical distances of the 
order of $1 \; \mathrm{Mpc}$. CHVCs would then be quite large and massive objects with sizes of about $15 \; \mathrm{kpc}$ 
and typical \ion{H}{i} masses of a few times $10^7 \; M_{\odot}$.

An improved CHVC catalogue with 67 objects on the basis of the LDS was published by de Heij et al. (\cite{deheij2}), using an 
automated search routine. In addition, Putman et al. (\cite{putman}) extracted 179 CHVCs from the \ion{H}{i} Parkes All-Sky Survey 
(HIPASS, Barnes et al. \cite{barnes}). Both catalogues for the northern and southern sky were then combined by de Heij et al. 
(\cite{deheij3}) for an all-sky catalogue of 216 CHVCs (30 clouds were located in the overlap region between LDS and HIPASS).

First doubts about a Local Group population of CHVCs arose with the non-detection of similar \ion{H}{i} clouds in other galaxy 
groups. Zwaan (\cite{zwaan}) investigated five galaxy groups with the Arecibo telescope to search for intragroup clouds. From 
his non-detection he concluded that the Local Group scenario proposed by Blitz et al. (\cite{blitz}) can be ruled out with 
$> 99$\% confidence level if the Local Group is typical among the investigated groups. A similar survey was conducted by 
Pisano et al. (\cite{pisano}) who studied three galaxy groups with the Parkes 64-m telescope and the Australia Telescope 
Compact Array (ATCA). They were not successful in detecting any \ion{H}{i} clouds similar to CHVCs, concluding that the known population 
of CHVCs has to be located within $160 \; \mathrm{kpc}$ from the Milky Way.

Further support for a circumgalactic population of CHVCs comes from Br\"uns et al. (\cite{bruens}) who discovered a 
pronounced head-tail structure of \object{CHVC 125$+$41}, suggesting that gas had been stripped off the cloud by ram-pressure interaction 
with the surrounding medium. This suggests a location of the cloud in the vicinity of the Milky Way. By applying the virial theorem 
to a compact clump within the cloud, Br\"uns et al. (\cite{bruens}) constrained the distance of \object{CHVC 125$+$41} to $d \approx 130 \; 
\mathrm{kpc}$ (in the case of no additional molecular gas) which is consistent with a circumgalactic population of CHVCs. Head-tail 
structures are also observed towards most of the large HVC complexes. Br\"uns et al. (\cite{bruens2}) systematically searched 
the LDS for distinct HVCs with $N_{\ion{H}{i}} \ge 10^{19} \; \mathrm{cm^{-2}}$. They found that about 20\% of the 252 clouds 
in their catalogue show a head-tail structure suggesting the presence of ram-pressure interaction between these HVCs and the 
ambient medium.

\begin{figure*}
\centering
  \includegraphics[width=\linewidth]{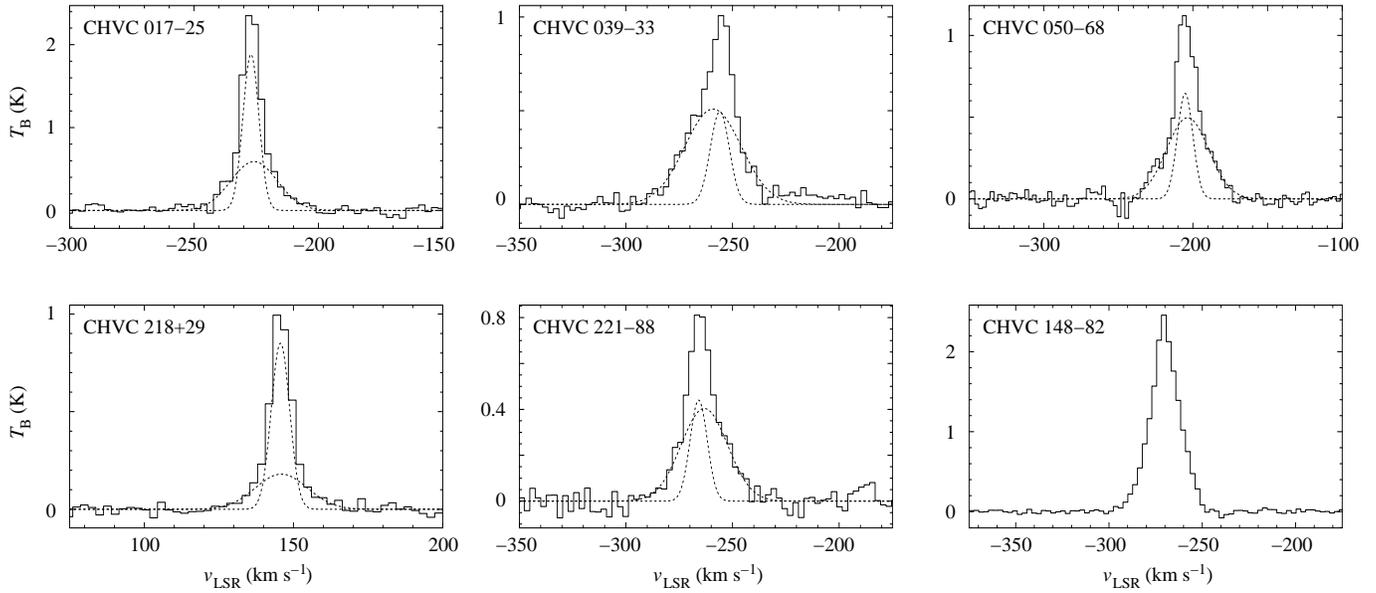}
  \caption{Example spectra of the five CHVCs with two-component line profiles and of \object{CHVC 148$-$82}. The dotted curves 
    represent a decomposition of the spectra into two Gaussian components. The superposition of a broad and a narrow Gaussian 
    component indicates the presence of two more or less distinct phases of warm and cold neutral gas. In the case of \object{CHVC 148$-$82} 
    a separation into two Gaussian components is not useful, although the narrow cusp on top of the \ion{H}{i} lines 
    towards the centre of the cloud again indicates the existence of a compact core of cold neutral gas.}
  \label{fig_spectra_twogauss}
\end{figure*}

A circumgalactic CHVC population is also supported by hydrodynamical simulations of Sternberg et al. (\cite{sternberg}) who 
considered CHVCs as spherically-symmetric, pressure-supported gas clouds confined by a gravitationally dominant dark-matter halo. 
They simulated both a Local Group population of CHVCs with a typical distance of $750 \; \mathrm{kpc}$ and a circumgalactic 
distribution with a distance of $150 \; \mathrm{kpc}$. Sternberg et al. (\cite{sternberg}) compared their simulations 
with the observed properties of CHVCs according to high-resolution observations with the Westerbork Synthesis Radio Telescope (WSRT) 
by Braun \& Burton (\cite{braun2}) and Arecibo observations by Burton et al. (\cite{burton}). They found the observed properties to 
be consistent with the results of their circumgalactic model so that CHVCs might be associated with an extended Galactic corona. 
At a typical distance of $150 \; \mathrm{kpc}$ the average \ion{H}{i} mass of CHVCs would be $M_{\ion{H}{i}} \approx 3 \cdot 10^5 
\; M_{\odot}$. Furthermore, a diffuse Galactic corona gas can provide the external pressure required to additionally support CHVCs 
and to account for head-tail structures like the one observed in the case of \object{CHVC 125$+$41}.

The results of Sternberg et al. (\cite{sternberg}) also outline observational strategies to distinguish between a Local Group 
origin of CHVCs and a circumgalactic CHVC population. In the circumgalactic case, their simulations predict a sharp drop in the 
\ion{H}{i} column density below $N_{\ion{H}{i}} \approx 5 \cdot 10^{18} \; \mathrm{cm^{-2}}$ due to ionisation by the strong 
intergalactic radiation field in the vicinity of the Galaxy. In addition, the circumgalactic model predicts possible gas wings 
with \ion{H}{i} column densities below $10^{18} \; \mathrm{cm^{-2}}$. In the Local Group scenario the \ion{H}{i} colum density 
shows a smooth decline at the edge of the clouds and no extended wings are expected. Sensitive observations with the $9'$ HPBW 
of the 100-m telescope in Effelsberg should, thus, allow us to evaluate the distribution of CHVCs by investigating the column 
density profile across a sample of objects in detail. Another aim of our observations is the search for head-tail structures 
among CHVCs which would suggest the presence of an ambient medium and could provide an additional constraint on the distance 
and distribution of CHVCs.

This paper is organised as follows: In Sect.~\ref{sect_sample} we describe how the 11 investigated CHVCs were chosen. 
Sect.~\ref{sect_data} explains the entire data acquisition and reduction process. In Sect.~\ref{sect_results} we present 
the results of our survey and summarise the physical parameters of our 11~CHVCs. In Sect.~\ref{sect_discussion} we 
discuss the implications of our results for the origin and the properties of CHVCs. Sect.~\ref{sect_summary} 
summarises our results and conclusions.

\section{\label{sect_sample}Sample selection}

10 of the 11 CHVCs were selected from the Braun \& Burton (\cite{braun}) catalogue on the basis of previous, less sensitive 
Effels\-berg \ion{H}{i} observations of 41 CHVCs (Westmeier \cite{westmeier}) by two criteria. First, all clouds still had to be 
classified as CHVCs in the improved catalogue by de Heij et al. (\cite{deheij2}). Second, we selected only those clouds with an 
\ion{H}{i} peak column density ratio of $N_{\mathrm{\ion{H}{i}}}^{\mathrm{Eff}} / N_{\mathrm{\ion{H}{i}}}^{\mathrm{LDS}} \ge 3$ 
between the Effelsberg data and the LDS data. A high column density ratio implies that the clouds contain 
compact substructure which is unresolved with the 25-m Dwingeloo telescope ($36'$ HPBW). The obtained column density ratios are 
all well below the value expected for point sources so that our 11 CHVCs must all be partly resolved with the 100-m Effelsberg 
telescope ($9'$ HPBW). Thus, the latter criterion selects the most compact clouds for re-investigation which are not completely 
resolved with the existing LDS data and worth being re-observed with the higher angular resolution of the 100-m telescope.

The remaining object, \object{CHVC 218$+$29}, was selected from the de Heij et al. (\cite{deheij2}) catalogue to obtain a complete coverage 
of the available sidereal time interval. \object{CHVC 218$+$29} was attractive because of its northern latitude, its small angular 
size and its unusually narrow \ion{H}{i} lines of only $7 \; \mathrm{km \, s^{-1}}$ FWHM according to the de Heij et al. 
(\cite{deheij2}) catalogue.

\begin{figure*}
\centering
  \includegraphics[width=17cm]{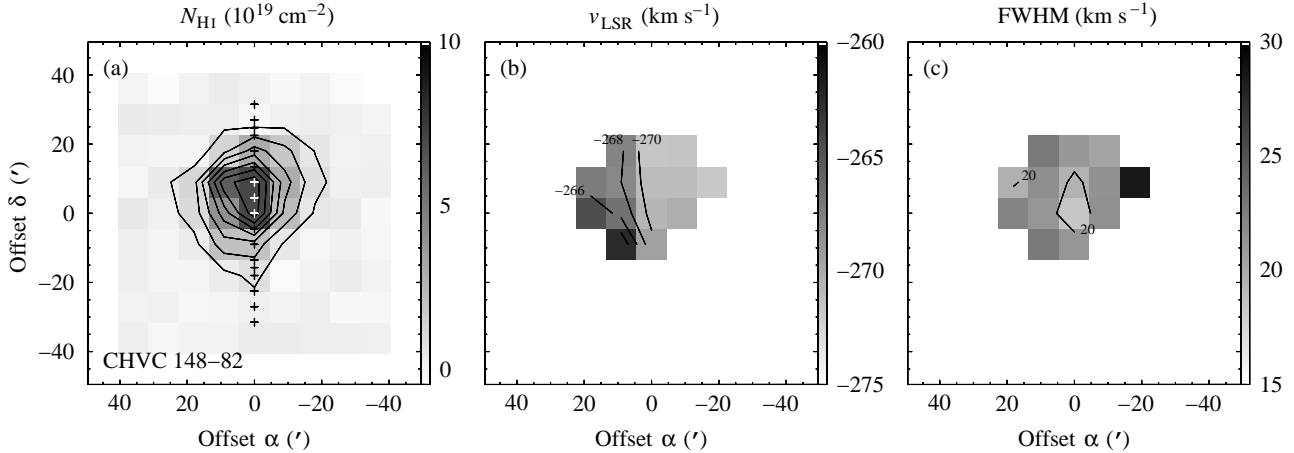}
  \caption{\object{CHVC 148$-$82} as described in Sect.~\ref{sect_ss}. \textbf{(a)} Integrated \ion{H}{i} column density map. Contours 
    start at $5 \cdot 10^{18} \; \mathrm{cm^{-2}}$ with an increment of $1 \cdot 10^{19} \; \mathrm{cm^{-2}}$. The crosses indicate 
    the positions of individual spectra along the deep slice (see Fig.~\ref{fig_cut_ss}). \textbf{(b)} Distribution of LSR radial 
    velocities of the gas. \textbf{(c)} Distribution of line widths (FWHM).}
  \label{fig_overview_ss}
\end{figure*}

\section{\label{sect_data}Data acquisition and reduction}

A map of $11 \times 11$ spectra on a $9'$ grid (beam-by-beam sampling) was observed for most clouds, resulting in 
a field of view of $1.5^{\circ} \times 1.5^{\circ}$. The central position was derived from previous, less sensitive Effelsberg 
observations (Westmeier \cite{westmeier}) so that the object should be centred in the map. Each \ion{H}{i} spectrum was 
integrated for 3~minutes in the in-band frequency switching mode, using the 1024-channel autocorrelator. The bandwidth was 
$6.3 \; \mathrm{MHz}$, resulting in a velocity resolution of $2.6 \; \mathrm{km \, s^{-1}}$. Signal and reference spectra were 
shifted by $\pm 1.4 \; \mathrm{MHz}$ relative to the central frequency. The advantage of this in-band frequency switching method 
is that the \ion{H}{i} line of the CHVC is included in both the signal and reference spectra so that the entire integration time 
is used for detection of the astronomical signal, resulting in a typical baseline rms of $\sigma_{\mathrm{rms}} \approx 50 \; 
\mathrm{mK}$ at the original velocity resolution of $2.6 \; \mathrm{km \, s^{-1}}$. The maps allow us to determine the overall 
morphology of the CHVCs and to identify possible head-tail structures or other signs of ram-pressure distortion.

Furthermore, we intended to compare the column density profiles of CHVCs with the predictions of the numerical simulations 
by Sternberg et al. (\cite{sternberg}). For this purpose, we observed additional positions along the apparent symmetry axis of 
each CHVC with a longer integration time of typically 10~minutes per spectrum and a denser angular sampling of $4.5'$ or $6.4'$ 
between two adjacent spectra. These deep slices allow us to investigate the \ion{H}{i} column density profile in more detail 
and with higher sensitivity. The typical baseline rms along the deep slices is $\sigma_{\mathrm{rms}} \approx 25 \ldots 35 \; 
\mathrm{mK}$ at the original $2.6 \; \mathrm{km \, s^{-1}}$ velocity resolution. As the deep slices in most cases had to be 
observed prior to a detailed analysis of the maps, there are two cases in which the slices were unfortunately not perfectly 
placed along the symmetry axis of the corresponding clouds.

In the case of \object{CHVC 218$+$29} we used a slightly different procedure. The object turned out to be very compact and elongated so 
that we chose a map size of $15 \times 9$ spectra on a $4.5'$ grid, corresponding to a field of view of roughly $70' \times 40'$. 
Furthermore, some \ion{H}{i} lines of \object{CHVC 218$+$29} turned out to be so narrow ($\approx 4 \; \mathrm{km \, s^{-1}}$ FWHM) that 
the velocity resolution of $2.6 \; \mathrm{km \, s^{-1}}$ could not sufficiently resolve them. Therefore, we had to observe \object{CHVC 
218$+$29} with a smaller bandwidth of $3.2 \; \mathrm{MHz}$, leading to a velocity resolution of $1.3 \; \mathrm{km \, s^{-1}}$. 
This made the use of the in-band frequency switching method inappropriate so that we had to observe \object{CHVC 218$+$29} in the normal frequency 
switching mode, resulting in a slightly higher baseline rms of $\sigma_{\mathrm{rms}} = 88 \; \mathrm{mK}$ at the original 
$1.3 \; \mathrm{km \, s^{-1}}$ velocity resolution.

All spectra were calibrated using the S7 standard calibration source (Kalberla et al. \cite{kalberla2}). The statistical 
uncertainties in the derived calibration factors are in the range of $1 \ldots 3$\% for each observing session. Next, the 
standard stray radiation correction was applied to the spectra (Kalberla et al. \cite{kalberla}). Because of their high radial 
velocities most CHVCs are only marginally affected by Galactic stray radiation. However, a few of our CHVCs have radial velocities 
of $|v_{\mathrm{LSR}}| \lesssim 200 \; \mathrm{km \, s^{-1}}$ which is in the regime of Galactic \ion{H}{i} gas towards some 
directions so that a stray radiation correction was in general recommended and performed for all spectra. After calibration and 
stray radiation correction all spectra were available in the form of brightness temperature $T_{\mathrm{B}}$ versus radial 
velocity in the local standard-of-rest frame $v_{\mathrm{LSR}}$.

The baseline correction for each spectrum was done by visual inspection, using the GILDAS package. As a result of the 
relatively narrow spectral lines of the CHVCs, the line windows could be set visually and a polynomial -- usually of order 3 
-- was then subtracted from the spectra. To obtain the physical parameters of our CHVCs (e.g. \ion{H}{i} column density, line width 
and radial velocity of the gas) a single Gaussian was fit to each spectrum. Finally, only those \ion{H}{i} lines with a brightness 
temperature of $T_{\mathrm{B}} \ge 3 \sigma_{\mathrm{rms}}$ were selected for further analysis and all other spectra were rejected. 
Unless otherwise noted, all maps and slices presented in this paper are based on the results of these Gaussian fits. Only the column 
density maps were derived from the zeroth moment of the spectra under the general assumption that the optical depth of the gas 
is negligible. In some cases, the \ion{H}{i} lines disclose a two-component structure indicating the presence of a cold and a warm 
gas phase. If so, we additionally fitted two Gaussian components to the spectra in order to study the properties of the two 
gas phases separately. Some examples of two-component line profiles are shown in Fig.~\ref{fig_spectra_twogauss}.

\begin{figure*}
\centering
  \includegraphics[width=\linewidth]{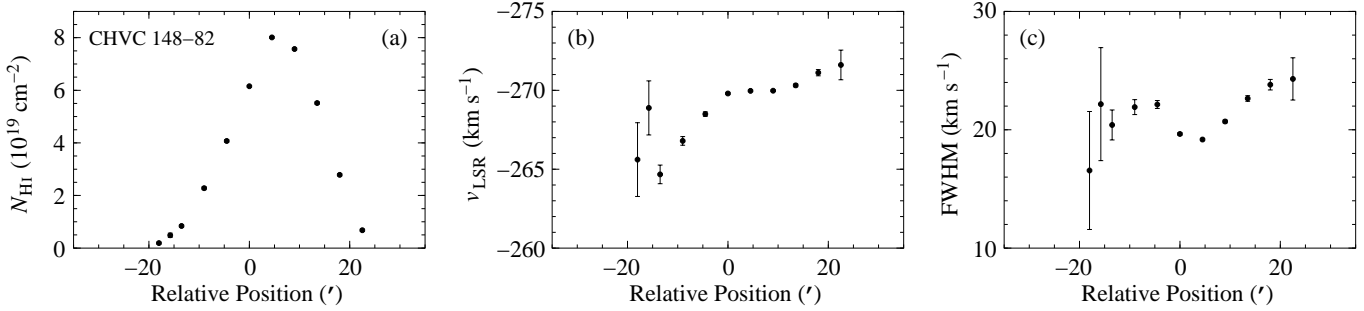}
  \caption{\object{CHVC 148$-$82} as described in Sect.~\ref{sect_ss}. \textbf{(a)} Integrated \ion{H}{i} column density along the slice 
    indicated by the crosses in Fig.~\ref{fig_overview_ss}~(a). \textbf{(b)} Distribution of LSR radial velocities of the gas along 
    the slice. \textbf{(c)} Distribution of line widths (FWHM) along the slice.}
  \label{fig_cut_ss}
\end{figure*}

\section{\label{sect_results}Results}

In this section we describe the physical parameters of the observed CHVCs. Sect.~\ref{sect_overall_properties} gives a general 
overview of the properties of our CHVC sample. Sects.~\ref{sect_ss} to \ref{sect_ir} address the properties of the 
individual clouds in more detail.

\subsection{\label{sect_overall_properties}Overall properties of the 11 CHVCs}

One motivation for our observations was the search for head-tail structures like the ones found in the case of \object{CHVC 125$+$41} 
(Br\"uns et al. \cite{bruens}) and in many HVC complexes (Br\"uns et al. \cite{bruens2}). It turned out that only one object, \object{CHVC 148$-$82}, 
has a spherically-symmetric appearance. All other CHVCs exhibit a more or less complex and irregular morphology. We identified 
6~CHVCs with either a head-tail structure or a bow-shock shape which in many cases is suggestive of ram-pressure interaction 
between the CHVCs and the ambient medium. The remaining clouds appear irregular in the sense that they do not exhibit a 
simple radial or axial symmetry. In Sects.~\ref{sect_ss} to \ref{sect_ir} we describe the properties of the individual CHVCs divided 
into the four morphological categories mentioned above. This morphological classification, however, is subjective and not 
based on a quantitative analysis. Nonetheless, we applied it to structure the description of the individual CHVCs with the objective 
of improving the readability of the paper. Our attempt to quantify the different morphologies on the basis of ellipse fits was not successful 
because of the relatively poor sampling of the CHVCs even with the 100-m telescope. A single pixel could significantly influence the 
fits and, thus, the morphological classification of a cloud.

The derived physical parameters of the 11 investigated CHVCs are summarised in Table~\ref{chvc_table}. With the exception of 
\object{CHVC 218$+$29} the radial velocities of the clouds are all negative in both the local standard-of-rest frame (LSR) and the 
Galactic standard-of-rest frame (GSR). This is simply based on a selection effect: The motion of Galactic rotation is superposed 
on the radial velocities of the CHVCs so that CHVCs in the northern sky have predominantly negative radial velocities in the LSR 
frame while CHVCs in the southern hemisphere have mainly positive radial velocities (also see Braun \& Burton \cite{braun}). We can 
account for the contribution of Galactic rotation by converting the radial velocities into the GSR frame. The asymmetry between positive 
and negative velocities, however, remains so that even in the GSR frame CHVCs in the northern hemisphere must have a net negative 
radial velocity. This means that the net radial velocity of CHVCs is of no use if we do not consider a homogeneous 
all-sky survey. Even in the latter case gas clouds of different origin (e.g. fragments of the Magellanic Stream or gas clouds 
of Galactic origin) might significantly influence the obtained mean radial velocity of the CHVC population so that the 
implications of the mean radial velocity of CHVCs must be considered with great care.

There is a wide range of different velocity gradients across the 11 investigated CHVCs. The two extreme cases are \object{CHVC 218$+$29} 
which shows velocity variations of only $4 \; \mathrm{km \, s^{-1}}$ and \object{CHVC 358$+$12} with a total gradient of about $85 \; 
\mathrm{km \, s^{-1}}$. In a few cases, rapid velocity changes seem to be induced by the presence of multiple gas components 
along the line of sight with different radial velocities. This is likely the case for \object{CHVC 050$-$68}, where we observe a velocity 
gradient of $60 \; \mathrm{km \, s^{-1}}$ across an angular distance of only $20'$. Broad and complex line profiles in this 
area suggest that the gradient is induced by individual gas components along the line of sight. In 
other cases, velocity variations are fairly regular with Gaussian line profiles, so that the velocity gradients 
can be interpreted as a sign of rotation of the clouds. This is in particular the case for \object{CHVC 358$+$12} where the velocity 
profile along the major axis of the cloud resembles the rotation curve of a galaxy.

The average line width of the investigated CHVCs is $23 \pm 7 \; \mathrm{km \, s^{-1}}$~FWHM. The narrowest lines by far are 
found in the case of \object{CHVC 218$+$29}, measuring around $4 \; \mathrm{km \, s^{-1}}$~FWHM at the eastern edge of the cloud. This 
corresponds to an upper limit in kinetic gas temperature of about $350 \; \mathrm{K}$.\footnote{$T = m_{\mathrm{H}} \Delta v^2 
/ (8 k \ln 2)$ according to the Maxwellian velocity distribution of an ideal gas where $m_{\mathrm{H}}$ is the mass of a 
hydrogen atom, $k$ the Boltzmann constant, and $\Delta v$ the FWHM of the \ion{H}{i} line.} The largest line widths of up to 
$60 \; \mathrm{km \, s^{-1}}$~FWHM are found in \object{CHVC 358$+$12}, but the complex line profiles in this case suggest that such 
large line widths are caused by the superposition of multiple gas components with different radial velocities. In some CHVCs, 
line profiles show a two-component structure. A narrow line component seems to be superposed on a broad component, indicating 
the presence of two phases of cold neutral medium (CNM) and warm neutral medium (WNM). Such two-component line profiles can be 
seen in \object{CHVC 017$-$25}, \object{CHVC 039$-$33}, \object{CHVC 050$-$68}, \object{CHVC 218$+$29} and \object{CHVC 221$-$88}. 
Example spectra of these clouds are shown in Fig~\ref{fig_spectra_twogauss}. In the case of \object{CHVC 148$-$82} (lower-right panel), 
no clear two-component structure is present in the spectra. The shape of the spectral lines, however, is not perfectly Gaussian in the 
central parts of the cloud. A narrow cusp on top of the \ion{H}{i} line indicates the presence of a compact cold core which is not 
resolved with the $9'$~HPBW of the 100-m telescope. Evidence for a cold core in \object{CHVC 148$-$82} was also found by de Heij et al. 
(\cite{deheij}). Their high-resolution synthesis observations with the WSRT show a very compact clump with \ion{H}{i} line widths of 
$\lesssim 10 \; \mathrm{km \, s^{-1}}$~FWHM.

\begin{figure*}
\centering
  \includegraphics[width=17cm]{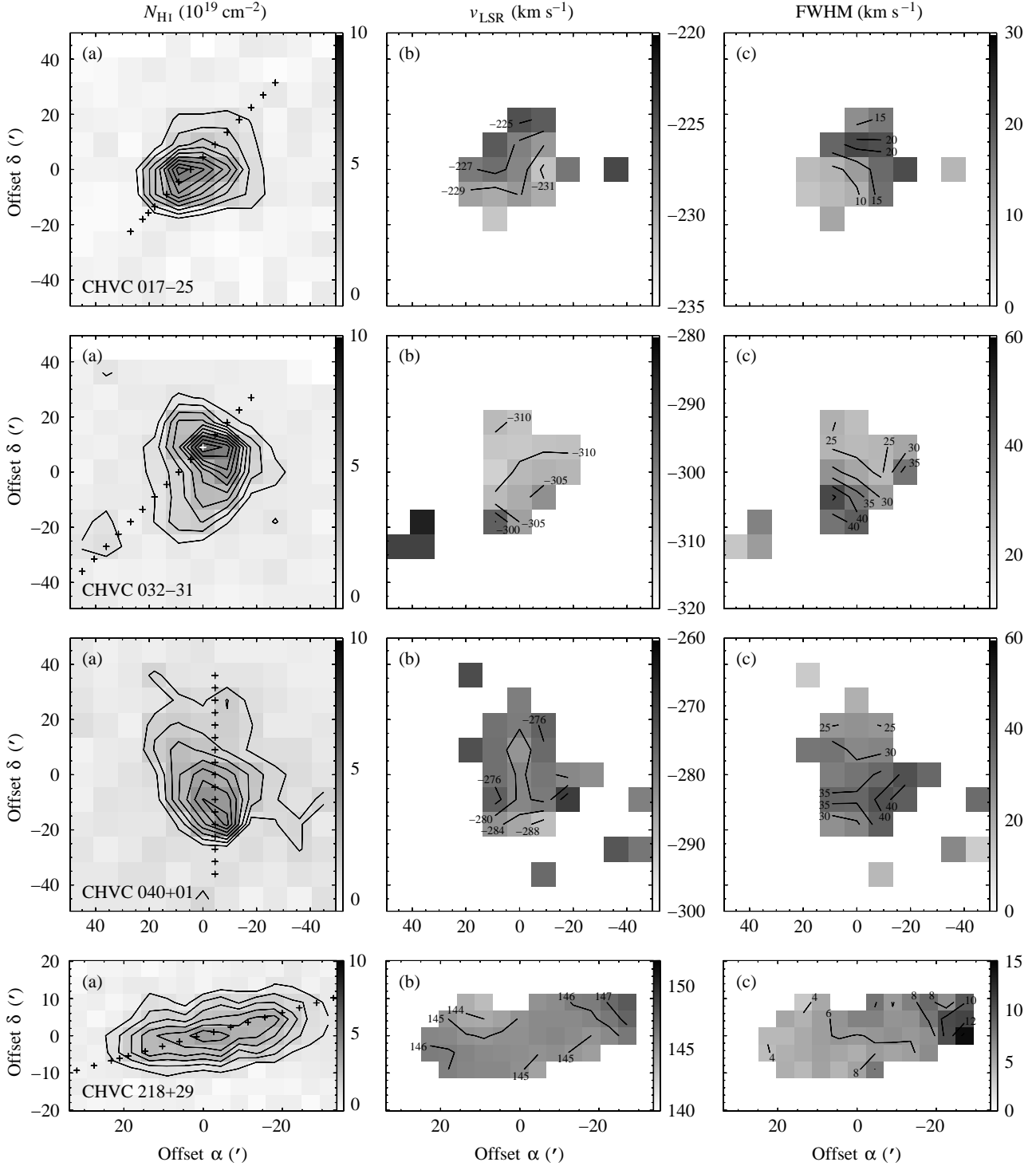}
  \caption{Head-tail CHVCs as described in Sect.~\ref{sect_ht}. \textbf{(a)} Integrated \ion{H}{i} column density map. Contours 
    start at $5 \cdot 10^{18} \; \mathrm{cm^{-2}}$ ($1 \cdot 10^{19} \; \mathrm{cm^{-2}}$ in the case of \object{CHVC 040$+$01}) with an 
    increment of $5 \cdot 10^{18} \; \mathrm{cm^{-2}}$. The crosses indicate the positions of individual spectra along the deep 
    slice (see Fig.~\ref{fig_cut_ht}). \textbf{(b)} Distribution of LSR radial velocities of the gas. \textbf{(c)} Distribution 
    of line widths (FWHM).}
  \label{fig_overview_ht}
\end{figure*}

\begin{figure*}
\centering
  \includegraphics[width=\linewidth]{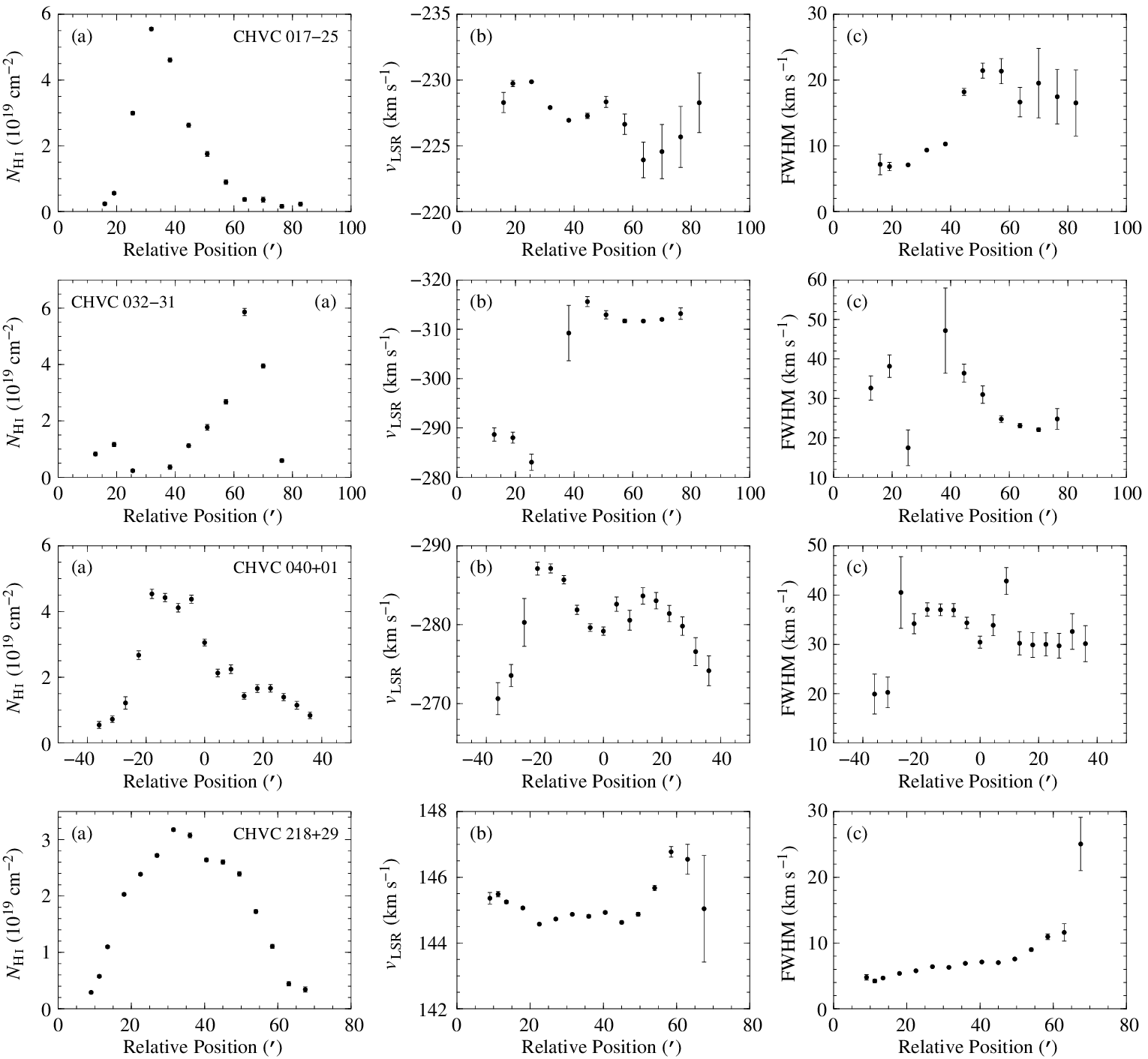}
  \caption{Head-tail CHVCs as described in Sect.~\ref{sect_ht}. \textbf{(a)} Integrated \ion{H}{i} column density along the 
    slice indicated by the crosses in Fig.~\ref{fig_overview_ht}~(a). \textbf{(b)} Distribution of LSR radial velocities of 
    the gas along the slice. \textbf{(c)} Distribution of line widths (FWHM) along the slice.}
  \label{fig_cut_ht}
\end{figure*}

\begin{figure*}
\centering
  \includegraphics[width=17cm]{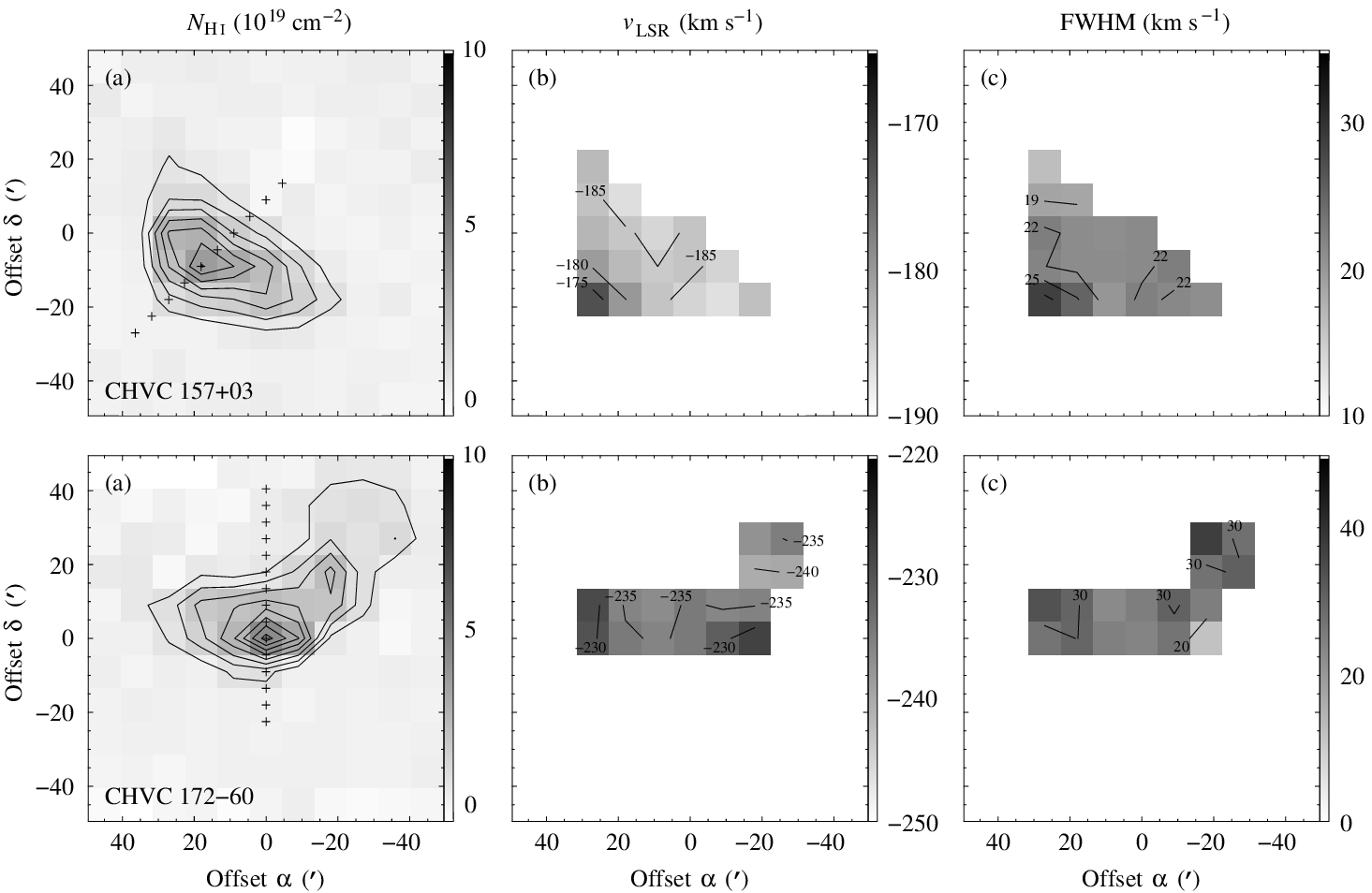}
  \caption{Bow-shock shaped CHVCs as described in Sect.~\ref{sect_bs}. \textbf{(a)} Integrated \ion{H}{i} column density map. 
    Contours start at $5 \cdot 10^{18} \; \mathrm{cm^{-2}}$ with an increment of $5 \cdot 10^{18} \; \mathrm{cm^{-2}}$. The 
    crosses indicate the positions of individual spectra along the deep slice (see Fig.~\ref{fig_cut_bs}). \textbf{(b)} 
    Distribution of LSR radial velocities of the gas. \textbf{(c)} Distribution of line widths (FWHM).}
  \label{fig_overview_bs}
\end{figure*}

\begin{figure*}
\centering
  \includegraphics[width=\linewidth]{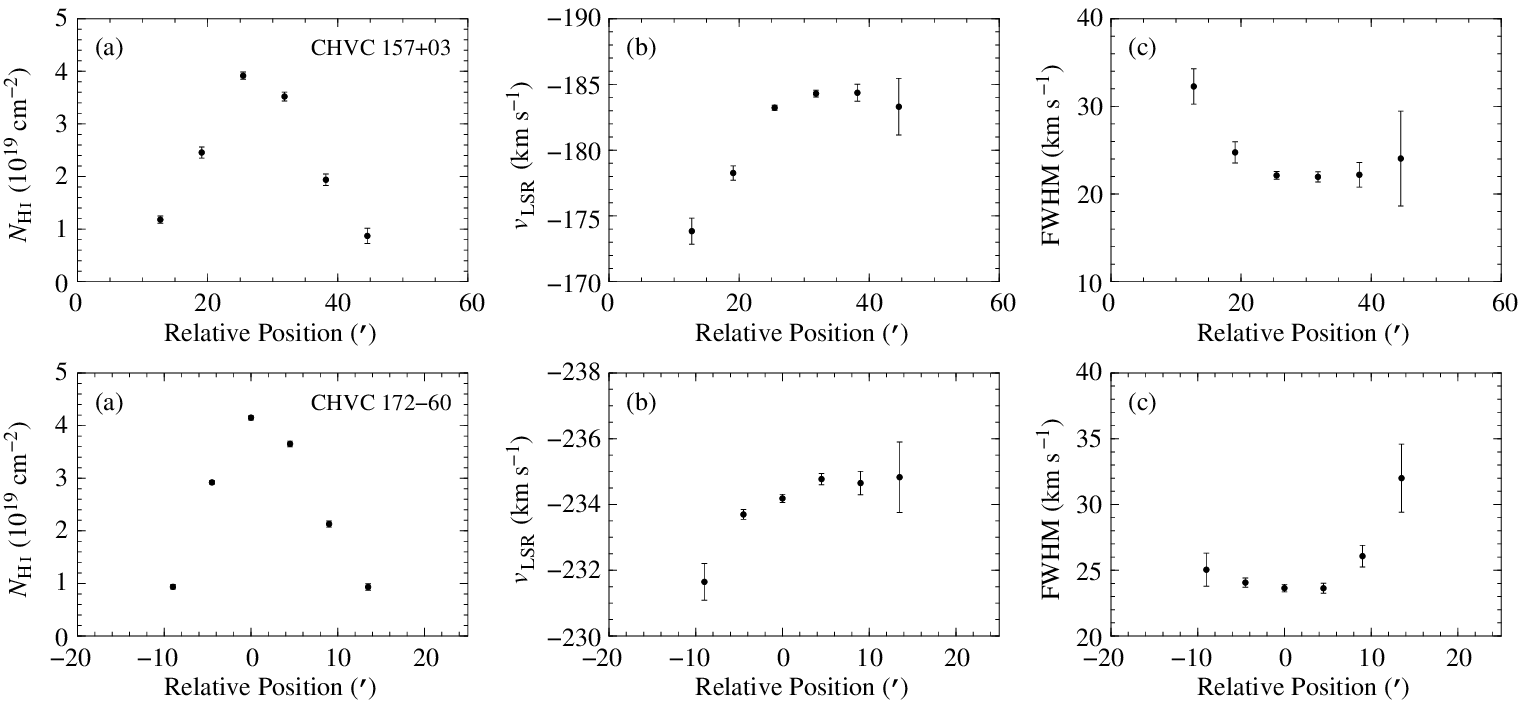}
  \caption{Bow-shock shaped CHVCs as described in Sect.~\ref{sect_bs}. \textbf{(a)} Integrated \ion{H}{i} column density along 
    the slice indicated by the black crosses in Fig.~\ref{fig_overview_bs}~(a). \textbf{(b)} Distribution of LSR radial 
    velocities of the gas along the slice. \textbf{(c)} Distribution of line widths (FWHM) along the slice.}
  \label{fig_cut_bs}
\end{figure*}

The average \ion{H}{i} peak column density of the CHVCs in the original catalogue by Braun \& Burton (\cite{braun}) is 
$N_{\mathrm{\ion{H}{i}}} = 2.1 \pm 1.6 \cdot 10^{19} \; \mathrm{cm}^{-2}$. The observed peak column densities of our 11 CHVCs 
are in the range of $N_{\mathrm{HI}} \approx 3 \ldots 8 \cdot 10^{19} \; \mathrm{cm}^{-2}$ with a slightly larger average 
value of $5.3 \pm 1.7 \cdot 10^{19} \; \mathrm{cm}^{-2}$. This is consistent with the value of $5 \cdot 10^{19} \; \mathrm{cm}^{-2}$ 
following from the circumgalactic model of Sternberg et al. (\cite{sternberg}). The peak brightness temperatures of the CHVCs in our 
survey range from $0.7 \; \mathrm{K}$ to $3.1 \; \mathrm{K}$ with an average of $1.6 \pm 0.8 \; \mathrm{K}$.

The ellipticities of the clouds were determined by a second moment analysis following the procedure described by Banks et al. 
(\cite{banks}). First, a clip level of $5 \cdot 10^{18} \; \mathrm{cm}^{-2}$ was applied to the column density map of each cloud, 
corresponding to about the 3-sigma detection limit of our data. Next, the central position of the cloud was derived from the column 
density weighted first moment. Then we calculated the column density weighted second moments for each map from which we obtained the 
major and minor axis, $a$ and $b$, of the ellipse as described by Banks et al. (\cite{banks}). From the major and minor axis we 
finally calculated the ellipticity $e$, using the definition $e = 1 - \frac{b}{a}$.

\subsection{\label{sect_ss}Spherically-symmetric CHVCs}

\subsubsection{\object{CHVC 148$-$82}}

\object{CHVC 148$-$82} is the only cloud among our 11 investigated CHVCs which has an almost spherically-symmetric appearance. 
It is located close to the Galactic south pole and the Magellanic Stream. The radial velocity of \object{CHVC 148$-$82}, however, is 
different by about $170 \; \mathrm{km \, s^{-1}}$ from the one observed for the Magellanic Stream ($v_{\mathrm{LSR}} \approx -100 
\; \mathrm{km \, s^{-1}}$, Br\"uns et al. \cite{bruens3}) in this area so that an association between both objects is unlikely. 
Fig.~\ref{fig_overview_ss} shows the distribution of \ion{H}{i} column densities, line widths and radial velocities. We find a 
clear velocity gradient across the cloud, measuring roughly $8 \; \mathrm{km \, s^{-1}}$ across an angular distance of about $45'$. 
The distribution of line widths is consistent with a spherical symmetry of \object{CHVC 148$-$82} as the \ion{H}{i} lines in the centre 
of the cloud are typically narrower than those at the edge. This is what we would expect for a spherically-symmetric gas cloud with 
decreasing gas temperature towards the centre.

The results of the deep and more detailed observations performed along the slice indicated by the crosses in 
Fig.~\ref{fig_overview_ss}~(a) are presented in Fig.~\ref{fig_cut_ss}.  The column density profile confirms the symmetric 
appearance of \object{CHVC 148$-$82} although towards the northern edge of the cloud the column density decreases more rapidly than 
towards the southern edge. This slight asymmetry is already indicated in the map in Fig.~\ref{fig_overview_ss}~(a). 
Unfortunately, the slice was not exactly chosen along the velocity gradient of \object{CHVC 148$-$82}. Nonetheless, the somewhat 
irregular velocity gradient is well traced by the data. The line width profile across the deep slice clearly shows the 
systematically smaller line widths in direction of the centre of \object{CHVC 148$-$82} which suggests the existence of cold gas in 
the centre of the cloud. Although a distinct cold gas component cannot directly be seen in the spectra, its existence is 
indicated by narrow cusps on top of the \ion{H}{i} lines in the direction of the cloud's centre (see Fig.~\ref{fig_spectra_twogauss}). 
These cusps are a direct hint of the existence of a compact core of cold neutral gas which is not resolved with the $9'$ HPBW 
of the Effelsberg telescope. This compact core was directly detected by de Heij et al. (\cite{deheij}) with 
high-resolution WSRT observations.

\subsection{\label{sect_ht}Head-tail CHVCs}

Four CHVCs from our sample exhibit a head-tail structure, suggesting the presence of ram-pressure 
interaction between these CHVCs and the ambient medium. Maps of the integrated column density and the distribution of radial 
velocities and line widths are shown in Fig.~\ref{fig_overview_ht}. The results of the deep and more detailed observations 
performed along the slice indicated by the crosses in Fig.~\ref{fig_overview_ht}~(a) are presented in Fig.~\ref{fig_cut_ht} 
as a function of relative position along the slice.

\subsubsection{\label{sect_chvc017-25}\object{CHVC 017$-$25}}

\object{CHVC 017$-$25} is located in the vicinity of the two HVC complexes GCP and GCN (Galactic Centre Positive/Negative). At first 
glance, the head-tail structure of \object{CHVC 017$-$25} does not seem very pronounced. The column density map shows that the spacings 
between contour lines are noticeably wider towards the north-western edge of the cloud, indicating the existence of a diffuse, 
faint gas tail in this direction. In the \ion{H}{i} column density distribution along the deep slice the head-tail structure 
of \object{CHVC 017$-$25} can clearly be seen. At the south-eastern edge there is a sharp increase in column density while towards the 
north-western edge we see a faint and extended tail. The behaviour of radial velocities is not completely regular; however, 
there is a general north-south gradient of about $5 \; \mathrm{km \, s^{-1}}$. Along the deep slice, radial velocities 
show a semi-periodic variation without any clear trend. The observed \ion{H}{i} line widths show a clear gradient with quite 
narrow lines in the south-eastern part of the cloud. The narrowest lines at the south-eastern edge measure only about 
$7 \; \mathrm{km \, s^{-1}}$ FWHM corresponding to an upper limit for the gas temperature of about $1100 \; \mathrm{K}$. 
In the direction of the extended tail of \object{CHVC 017$-$25} the line widths increase noticeably, reaching values around 
$20 \; \mathrm{km \, s^{-1}}$ FWHM which corresponds to an upper limit for the gas temperature of roughly $9000 \; \mathrm{K}$. 
These results suggest the separation of a cold and a warm gas component and substantiate the impression of a 
head-tail structure of \object{CHVC 017$-$25}. The obvious distortion of \object{CHVC 017$-$25} is discussed in more detail in 
Sect.~\ref{sect_ram-pressure}.

\subsubsection{\object{CHVC 032$-$31}}

\object{CHVC 032$-$31} can be found in the direction of the HVC complex GCN which combines several smaller clouds with similar negative 
radial velocities. The head-tail structure of \object{CHVC 032$-$31} is somewhat more pronounced than in the case of \object{CHVC 017$-$25}. At the 
north-western edge there is a sharp increase in column density while in the opposite direction the \ion{H}{i} emission is much 
more extended and ends in a faint tail. Another remarkable feature is the isolated emission in the lower left corner of the 
map. With a maximum \ion{H}{i} column density of $N_{\mathrm{\ion{H}{i}}} \approx 10^{19} \; \mathrm{cm}^{-2}$, this small 
cloud lies clearly above our detection limit. Within the significance of our data, it is not connected to the main body of \object{CHVC 
032$-$31}, resulting in an upper limit ($3 \, \sigma_{\mathrm{rms}}$) for the column density of a potential bridge of roughly 
$2.5 \cdot 10^{18} \; \mathrm{cm^{-2}}$ if a line width of $40 \; \mathrm{km \, s^{-1}}$ FWHM is assumed. The radial velocity 
of the small isolated cloud is significantly lower (w.r.t. the absolute value) than the velocities observed across the main 
cloud, indicating a separation in both position and velocity. The \ion{H}{i} lines across \object{CHVC 032$-$31} are comparatively broad 
with a mean value of $\langle \Delta v \rangle = 30 \pm 7 \; \mathrm{km \, s^{-1}}$. As in the case of \object{CHVC 017$-$25}, there is a 
strong line width gradient between head and tail.

\subsubsection{\object{CHVC 040$+$01}}

\object{CHVC 040$+$01} is located very close to the Galactic plane. From the morphological point of view the head-tail structure of 
\object{CHVC 040$+$01} is the most pronounced among our 11 CHVCs. The column density map shows a compact head and an extended tail in 
the north-eastern direction. This structure can also be seen in the column density profile along the deep slice although the 
slice unfortunately is not perfectly aligned with the symmetry axis of \object{CHVC 040$+$01}. Moreover, there is evidence of very 
faint and diffuse emission all across the map with intensities typically below $3 \, \sigma_{\mathrm{rms}}$. This faint 
emission is particularly prominent in the western part of the map where column densities of up to $N_{\mathrm{\ion{H}{i}}} 
\approx 1 \cdot 10^{19} \; \mathrm{cm}^{-2}$ can be found. The observed radial velocities across \object{CHVC 040$+$01} do not show 
a regular behaviour. There is no clear gradient, and radial velocities along the deep slice again show a semi-periodic 
variation as in the case of \object{CHVC 017$-$25}. The observed \ion{H}{i} line widths are quite large, reaching about $40 \; \mathrm{km 
\, s^{-1}}$ FWHM in the south-western part of the cloud. Unlike in all other head-tail CHVCs, the broadest lines can be 
found in the vicinity of the head of \object{CHVC 040$+$01}. Along the tail, line widths decrease to around $30 \; \mathrm{km \, 
s^{-1}}$ FWHM.

\begin{figure*}[!t]
\centering
  \includegraphics[width=17cm]{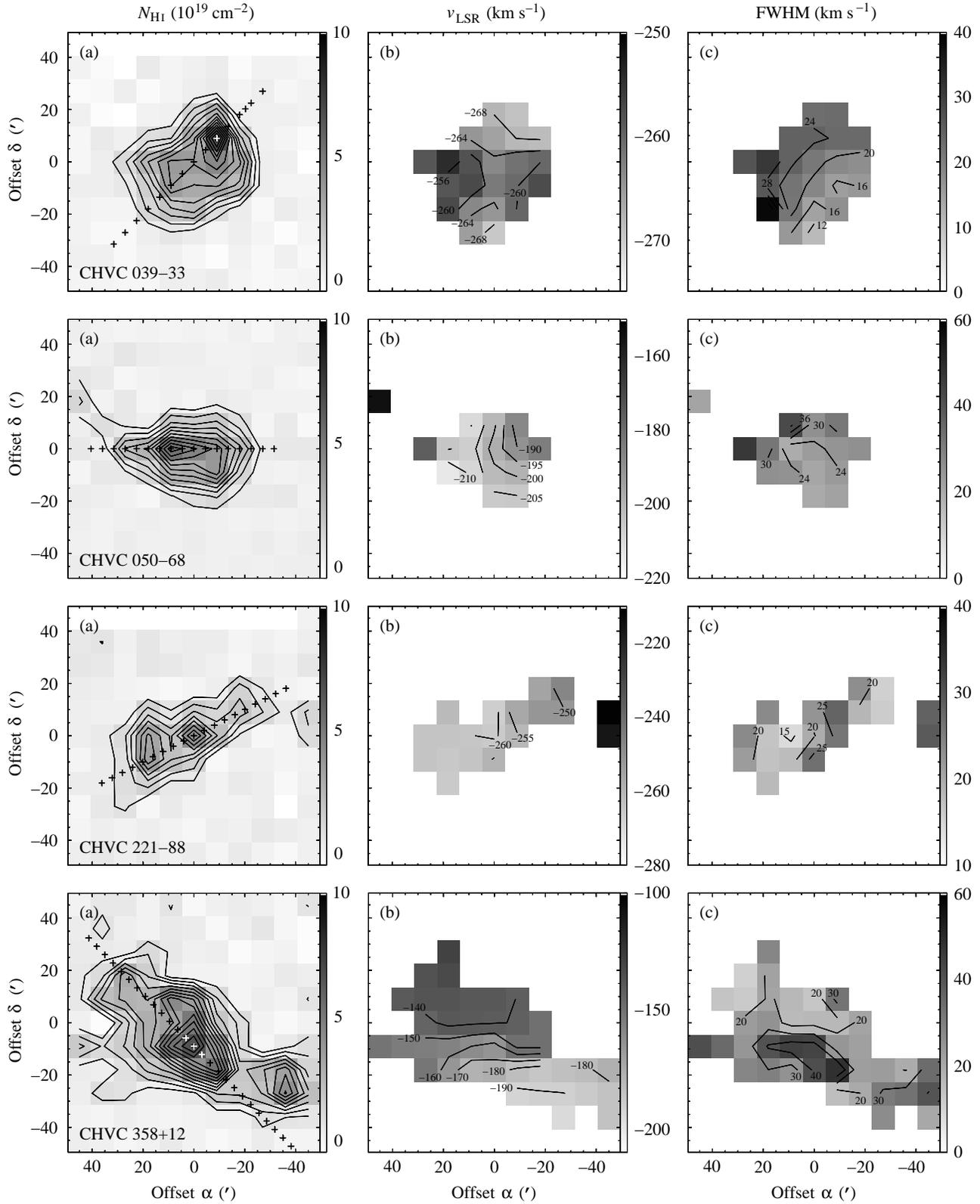}
  \caption{Irregular CHVCs as described in Sect.~\ref{sect_ir}. \textbf{(a)} Integrated \ion{H}{i} column density map. Contours 
    start at $5 \cdot 10^{18} \; \mathrm{cm^{-2}}$ with an increment of $5 \cdot 10^{18} \; \mathrm{cm^{-2}}$. The crosses 
    indicate the positions of individual spectra along the deep slice (see Fig.~\ref{fig_cut_ir}). \textbf{(b)} Distribution of 
    LSR radial velocities of the gas. \textbf{(c)} Distribution of line widths (FWHM).}
  \label{fig_overview_ir}
\end{figure*}

\begin{figure*}
\centering
  \includegraphics[width=\linewidth]{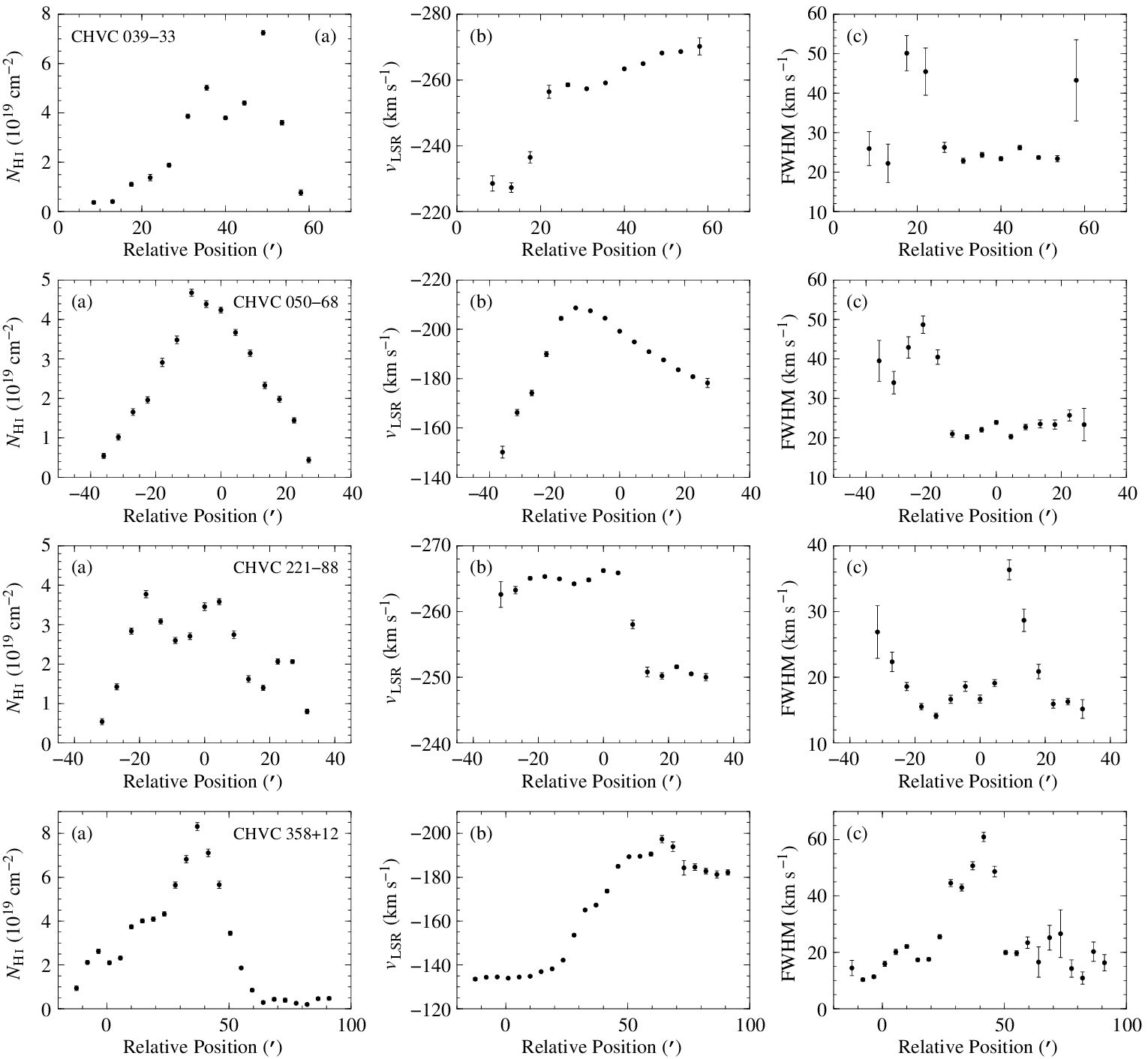}
  \caption{Irregular CHVCs as described in Sect.~\ref{sect_ir}. \textbf{(a)} Integrated \ion{H}{i} column density along the 
    slice indicated by the crosses in Fig.~\ref{fig_overview_ir}~(a). \textbf{(b)} Distribution of LSR radial velocities of 
    the gas along the slice. \textbf{(c)} Distribution of line widths (FWHM) along the slice.}
  \label{fig_cut_ir}
\end{figure*}

\subsubsection{\object{CHVC 218$+$29}}

\object{CHVC 218$+$29} is one of the most remarkable objects among our 11 CHVCs because it reveals some extreme properties. It is 
highly elongated with a major axis of about $1^{\circ}$ but a minor axis of only $20'$ with respect to the $5 \cdot 10^{18} 
\; \mathrm{cm}^{-2}$ column density level. The head-tail structure is not very prominent but the deep column density profile 
shows a steep slope on one side of the cloud and a more moderate slope on the other side, forming a short and diffuse tail. 
Radial velocities across the object are nearly constant with $v_{\mathrm{LSR}} \approx +145 \; \mathrm{km \, s^{-1}}$. Only 
towards the western edge is there a slight increase to about $+147 \; \mathrm{km \, s^{-1}}$. The detected line widths are the 
smallest by far among the CHVCs in our sample. At the eastern edge we observe line widths of only around $4 \; \mathrm{km \, 
s^{-1}}$ FWHM, corresponding to an upper limit for the gas temperature of only about $350 \; \mathrm{K}$. Moreover, we find a clear and 
systematic gradient along the major axis of \object{CHVC 218$+$29}. Towards the eastern edge, line widths rise above $10 \; \mathrm{km 
\, s^{-1}}$ FWHM. This prominent line width gradient again suggests the separation of a warm envelope and a cold gas component. 

\subsection{\label{sect_bs}Bow-shock shaped CHVCs}

Two CHVCs from our sample exhibit a bow-shock shape, suggesting the presence of ram-pressure interaction between 
these CHVCs and the ambient medium. Maps of the integrated column density and the distribution of radial velocities and line 
widths are shown in Fig.~\ref{fig_overview_bs}. The results of the deep and more detailed observations performed along the 
slice indicated by the crosses in Fig.~\ref{fig_overview_bs}~(a) are presented in Fig.~\ref{fig_cut_bs} as a function of 
relative position along the slice.

\subsubsection{\object{CHVC 157$+$03}}

\object{CHVC 157$+$03} is located close to the Galactic plane and is characterised by a slightly condensed core and two symmetric gas 
wings north-east and south-west of the core. It is a nice example of a CHVC with a bow-shock shape. Both radial 
velocities and line widths follow the axial symmetry of the column density distribution. Across most of \object{CHVC 157$+$03} the radial 
velocities of the gas are more or less the same with $v_{\mathrm{LSR}} \approx -185 \; \mathrm{km \, s^{-1}}$. Only towards 
the south-eastern edge do the radial velocities systematically decrease (w.r.t. their absolute values) by about $10 \; \mathrm{km \, 
s^{-1}}$ which is clearly traced by the velocity gradient along the deep slice. A similar behaviour is observed for the 
\ion{H}{i} line widths. Across most of the cloud the line widths are slightly above $20 \; \mathrm{km \, s^{-1}}$ FWHM, but towards 
the south-eastern edge they systematically rise above $30 \; \mathrm{km \, s^{-1}}$ FWHM.

\subsubsection{\object{CHVC 172$-$60}}

The appearance of \object{CHVC 172$-$60} is similar to that of \object{CHVC 157$+$03} although the core is more condensed and the wings are 
clearly asymmetric. The wing in the eastern direction is significantly less extended than the wing in the western direction. 
As in the case of \object{CHVC 157$+$03}, we observe a radial velocity gradient across the cloud which more or less follows the axial 
symmetry of the column density distribution. While we observe radial velocities around $-235 \; \mathrm{km \, s^{-1}}$ or 
higher across most of \object{CHVC 172$-$60}, velocities systematically decrease below $-230 \; \mathrm{km \, s^{-1}}$  towards the 
south-eastern and south-western flanks of the cloud. The behaviour of the \ion{H}{i} line widths, however, is different from 
that in \object{CHVC 157$+$03}. There is no clear increase in line width towards the southern edge of the cloud, but instead line widths 
grow noticeably towards the opposite side.

\subsection{\label{sect_ir}Irregular CHVCs}

The remaining four CHVCs from our survey show an irregular structure in the sense that their morphologies 
and the distribution of radial velocities and line widths do not exhibit a simple spherical or axial symmetry. Maps of 
the integrated column density and the distribution of radial velocities and line widths are shown in Fig.~\ref{fig_overview_ir}. 
The results of the deep and more detailed observations performed along the slice indicated by the crosses in 
Fig.~\ref{fig_overview_ir}~(a) are presented in Fig.~\ref{fig_cut_ir} as a function of relative position along the slice. 

\subsubsection{\object{CHVC 039$-$33}}

\object{CHVC 039$-$33} can be found near the HVC complex GCN. At first glance, \object{CHVC 039$-$33} appears to be 
a prime example of a head-tail CHVC. A very compact and condensed 
core is accompanied by an extended, diffuse gas tail in the south-eastern direction. The distribution of radial velocities and 
line widths, however, does not follow this axial symmetry. At both the northern and southern edge the highest radial velocities 
of $v_{\mathrm{LSR}} \approx -270 \; \mathrm{km \, s^{-1}}$ are observed, while the lowest velocities of $v_{\mathrm{LSR}} 
\gtrsim -260 \; \mathrm{km \, s^{-1}}$ can be found along the eastern and western edge of \object{CHVC 039$-$33}. This ``quadrupole'' 
symmetry of radial velocities does not fit into the simple picture of a head-tail CHVC. Concerning the distribution of \ion{H}{i} 
line widths, there is a strong gradient from south-west to north-east. In the south-western part of \object{CHVC 039$-$33} we observe 
relatively narrow lines with $\Delta v \approx 15 \; \mathrm{km \, s^{-1}}$ FWHM. Towards the north-eastern edge, the line widths 
reach $\Delta v \gtrsim 25 \; \mathrm{km \, s^{-1}}$ FWHM. Again, this clear gradient does not not follow the axial symmetry of 
the observed column density distribution but instead is perpendicular to the major axis of the cloud. The column density 
profile along the deep slice, which is shown in Fig.~\ref{fig_cut_ir}~(a), discloses a second maximum about $20'$ 
south-east of the compact core. This smaller peak is not visible in the map because it is too compact and is located exactly 
in between the individual pointings of the map. In addition, the deep column density profile affirms the impression of a 
head-tail structure of \object{CHVC 039$-$33}. At the north-western edge, there is a sharp increase in column density, while towards 
the south-eastern edge we detect a very faint and extended tail. But the structure of \object{CHVC 039$-$33} is more complicated than 
in the four cases of head-tail CHVCs discussed above. Therefore we classify it as an irregular CHVC.

\subsubsection{\object{CHVC 050$-$68}}

\object{CHVC 050$-$68} is located about $6^{\circ}$ away from the Magellanic Stream which has quite similar radial velocities in this 
area with $v_{\mathrm LSR} \approx -150 \ldots -200 \; \mathrm{km \, s^{-1}}$ (Br\"uns et al. \cite{bruens3}). Because of its 
extremely large radial velocity gradient \object{CHVC 050$-$68} is among the most remarkable CHVCs in our sample. Another conspicuous 
feature is a narrow, elongated arm in the north-eastern direction. This arm is rather faint, but at least one spectrum is clearly 
above our $3 \, \sigma_{\mathrm{rms}}$ detection limit. The most amazing feature of \object{CHVC 050$-$68}, however, is its velocity 
gradient. At the eastern edge, we find a radial velocity of $v_{\mathrm{LSR}} \approx -150 \; \mathrm{km \, s^{-1}}$. Towards 
the centre of the cloud, the radial velocity increases to $v_{\mathrm{LSR}} \approx -210 \; \mathrm{km \, s^{-1}}$ and then drops 
again to reach $v_{\mathrm{LSR}} \approx -180 \; \mathrm{km \, s^{-1}}$ at the western edge. This means that in the eastern part 
of \object{CHVC 050$-$68} we see a velocity gradient of $60 \; \mathrm{km \, s^{-1}}$ across only about $20'$. Across the western part 
of the cloud the velocity gradient still measures $30 \; \mathrm{km \, s^{-1}}$ across $40'$ and has the opposite sign. 
The somewhat smaller gradient in the western part of \object{CHVC 050$-$68} might indicate a rotation of the cloud. 
This is reasonable because the line widths observed in this part are almost constant. Moreover, the observed 
line profiles are regular and reveal a clear two-component structure, indicating the presence of two distinct cold and warm 
gas phases (see Fig.~\ref{fig_spectra_twogauss}). The eastern part of \object{CHVC 050$-$68}, however, is characterised by a sudden 
increase in line widths and complex, non-Gaussian line profiles. This indicates the presence of individual gas components 
with different radial velocities along the line of sight, whose superposition could give the impression of a strong velocity 
gradient across the eastern part of the cloud. This interpretation is supported by the extended, narrow arm in the north-eastern 
direction which exhibits radial velocities around $v_{\mathrm{LSR}} \approx -150 \; \mathrm{km \, s^{-1}}$, consistent with the 
values detected at the eastern edge of \object{CHVC 050$-$68}.

\subsubsection{\object{CHVC 221$-$88}}

\object{CHVC 221$-$88} is located near the Galactic south pole and about $5^{\circ}$ away from the Magellanic Stream. Nonetheless, a 
connection between both objects is unlikely because the Magellanic Stream reveals completely different radial velocities in this 
area with $v_{\mathrm{LSR}} \approx -50 \; \mathrm{km \, s^{-1}}$ (Br\"uns et al. \cite{bruens3}). \object{CHVC 221$-$88} is among the 
most complex CHVCs investigated in our survey. It is quite elongated with a major axis of about $1^{\circ}$ and a minor axis 
between $20'$ and $30'$. The column density map and, in more detail, the deep column density profile reveal three clumps, the 
westernmost of which is the faintest. Additional emission is detected at the western edge of the map. This emission has 
radial velocities around $v_{\mathrm{LSR}} \approx -210 \; \mathrm{km \, s^{-1}}$ which is about $40 \ldots 50 \; \mathrm{km 
\, s^{-1}}$ different from the velocities observed across the main part of \object{CHVC 221$-$88}. This means that despite their 
positional coincidence both clouds are kinematically clearly separated from each other. Unfortunately, we did not have enough 
time to extend our map in the western direction to investigate the second cloud in more detail. Focussing on the main cloud, 
its velocity structure is quite remarkable. While in the central and eastern parts of \object{CHVC 221$-$88} radial velocities range 
between $v_{\mathrm{LSR}} \approx -260 \ldots -265 \; \mathrm{km \, s^{-1}}$, they suddenly drop to $v_{\mathrm{LSR}} \approx 
-250 \; \mathrm{km \, s^{-1}}$ in the western part of the cloud. The velocity profile along the deep slice which is presented 
in Fig.~\ref{fig_cut_ir}~(b) shows that this drop by about $15 \; \mathrm{km \, s^{-1}}$ is extremely abrupt and occurs across 
only $9'$ (which is the HPBW of the Effelsberg telescope's main beam). In the intermediate region between both radial velocity 
regimes the observed line widths suddenly increase, indicating an increased velocity dispersion along the line of sight. At 
first glance, the velocity profile is reminiscent of a rotation curve. But the shape of the spectral lines in the intermediate region 
clearly exhibits two distinct gas components so that the two parts of \object{CHVC 221$-$88} with different radial velocities might be 
two separate clouds on the same line of sight rather than a single rotating cloud. The separate cloud at the western edge of 
the map supports this interpretation as it is also isolated in both velocity and position. It looks as if we see three individual 
clouds across the map with radial velocities decreasing (in absolute value) from east to west.

\subsubsection{\object{CHVC 358$+$12}}

\object{CHVC 358$+$12} is located about $12^{\circ}$ north of the Galactic centre. It is the largest among our 11 investigated 
CHVCs and extends beyond the boundaries of our map in both the eastern and western direction. \object{CHVC 358$+$12} reveals a rather 
complex structure with a compact core in the centre and several fainter clumps and extensions all across the object. The most 
remarkable feature, however, is the radial velocity gradient. In the south-western part of \object{CHVC 358$+$12} radial velocities 
up to $v_{\mathrm{LSR}} \approx -195 \; \mathrm{km \, s^{-1}}$ are observed. Towards the north-eastern edge the radial 
velocities drop to $v_{\mathrm{LSR}} \approx -130 \; \mathrm{km \, s^{-1}}$, and the deep spectra along the slice reveal a 
faint gas component which has even lower velocities of $v_{\mathrm{LSR}} \approx -110 \; \mathrm{km \, s^{-1}}$. This results 
in a total velocity gradient of $85 \; \mathrm{km \, s^{-1}}$ across the observed extent of the cloud. Furthermore, this velocity 
gradient resembles the flat rotation curve of a galaxy. This is already obvious in the velocity map in Fig.~\ref{fig_overview_ir}~(b) 
and is clearly illustrated by the velocity profile along the deep slice which was chosen along the velocity gradient and 
can be seen in Fig.~\ref{fig_cut_ir}~(b). This velocity gradient is accompanied by multiple line components especially in the 
central parts of \object{CHVC 358$+$12}. The occurrence of such multiple components is traced by larger line widths of up to $60 \; 
\mathrm{km \, s^{-1}}$ FWHM near the centre of the cloud. The properties of this remarkable CHVC will be discussed in more 
detail in a subsequent paper.

\begin{figure*}
\centering
  \includegraphics[width=16cm]{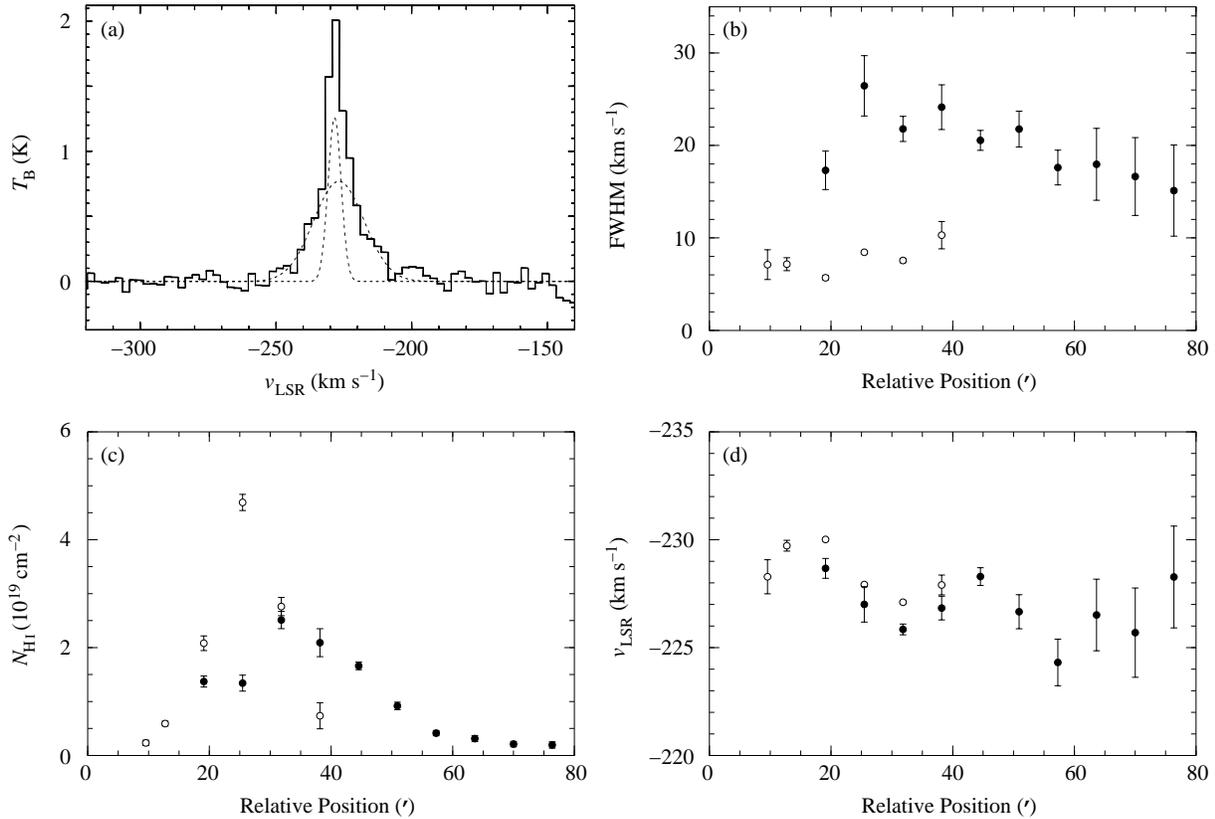}
  \caption{\object{CHVC 017$-$25}. \textbf{(a)} Typical spectrum near the centre of the cloud which shows a two-component line profile. 
    The dotted curves represent the decomposition of the spectrum into two Gaussian components. The three diagrams show the 
    distribution of \textbf{(b)} line widths, \textbf{(c)} \ion{H}{i} column densities and \textbf{(d)} LSR radial velocities 
    along the deep slice for the cold ($\circ$) and warm ($\bullet$) gas component. The error bars indicate the statistical 
    uncertainties of the Gaussian fit.}
  \label{fig_chvc01_twogauss}
\end{figure*}

\section{\label{sect_discussion}Discussion}

\subsection{\label{sect_ram-pressure2}Evidence of ram-pressure interaction}

The most outstanding result of our observations is the morphological complexity of the investigated CHVCs. Ten out of eleven 
clouds reveal an obviously non-spherical morphology with four CHVCs appearing irregular. When taking spectra along 
an appropriate axis across each cloud, we originally intended to obtain sensitive and detailed column density profiles at the 
clouds' edges and to compare our observations with the results of the hydrodynamical simulations by Sternberg et al. (\cite{sternberg}). 
But the explicitly irregular morphologies of most of the 11 observed CHVCs make a comparison of column density profiles and 
scale lengths nearly impossible. The simulations by Sternberg et al. (\cite{sternberg}) were based on the assumption of 
spherically-symmetric and undisturbed gas clouds. In many cases, however, we see an obvious asymmetry in the \ion{H}{i} 
distribution which may be an indication of interaction effects between the CHVCs and an ambient medium.

The complex morphologies of the CHVCs in our sample are not a selection effect caused by the selection criteria discussed in 
Sect.~\ref{sect_sample}. Virtually all of the 41 CHVCs in our previous, less sensitive Effelsberg survey (Westmeier 
\cite{westmeier}) disclose a non-spherical morphology, although the data were not sensitive enough to trace the distribution 
of the faint outer parts of the WNM. Thus, a complex, non-spherical distribution of \ion{H}{i} column densities seems to be 
typical for the known population of CHVCs, and spherically-symmetric clouds like \object{CHVC 148$-$82} are a rare exception.

The effects of ram-pressure stripping of HVCs were investigated by Quilis \& Moore (\cite{quilis}) using three-dimensional 
hydrodynamical simulations. They present pure gas clouds as well as dark-matter confined clouds in an ambient medium with 
a variety of different gas densities. They find that gas tails with \ion{H}{i} column densities $\gtrsim 10^{19} \; 
\mathrm{cm}^{-2}$ appear when the density of the ambient medium exceeds about $10^{-4} \; \mathrm{cm}^{-3}$. Even with lower 
gas densities around $10^{-5} \; \mathrm{cm}^{-3}$ faint gas tails emerge in their simulations with \ion{H}{i} column densities 
of about $10^{18} \; \mathrm{cm}^{-2}$.

The most prominent candidates for interacting clouds in our sample are the four head-tail CHVCs (see Fig. \ref{fig_overview_ht} 
and \ref{fig_cut_ht}). The head-tail morphology of these objects already suggests the presence of ram pressure. This impression 
is additionally supported by the distribution of line widths. With the exception of \object{CHVC 040$+$01}, all head-tail CHVCs reveal 
a clear line width gradient with the narrowest \ion{H}{i} lines appearing close to the head at the apparent leading edge of 
the cloud. This indicates a spatial separation of warm and cold neutral gas. In the case of \object{CHVC 218$+$29}, the bare cold gas lies 
exposed to its environment on one edge of the cloud with line widths around $4 \; \mathrm{km \, s^{-1}}$ FWHM. A similar 
situation is observed for \object{CHVC 017$-$25}, where line widths of about $7 \; \mathrm{km \, s^{-1}}$ FWHM are measured at the apparent 
leading edge.

Ram-pressure distortion is also a possible explanation for our two cases of bow-shock shaped CHVCs (see Fig. 
\ref{fig_overview_bs} and \ref{fig_cut_bs}). A detailed analysis of the distribution of radial velocities of the \ion{H}{i} 
gas reveals a clear velocity gradient across both clouds. While radial velocities are more or less constant across most parts 
of the two CHVCs, they decrease significantly towards the presumed leading edges. This suggests a deceleration of the gas due 
to the ram pressure of the ambient medium. In the case of \object{CHVC 157$+$03}, this velocity gradient measures about 
$10 \; \mathrm{km \, s^{-1}}$ and is accompanied by a clear line width gradient with opposite sign. Line widths systematically 
increase towards the apparent leading edge which may indicate heating or growing turbulence of the gas in the course of 
ram-pressure interaction.

Burton et al. (\cite{burton}) also reported significant asymmetries in the high column density regime ($N_{\mathrm{\ion{H}{i}}} 
> 3 \cdot 10^{18} \; \mathrm{cm}^{-2}$) of their 10 CHVCs observed with the Arecibo telescope. In the low column density regime below about 
$3 \cdot 10^{18} \; \mathrm{cm}^{-2}$, however, they found a high degree of reflection symmetry which they took as evidence against 
an external mechanism producing the observed asymmetries in the high column density cores. But their analysis is based only on 
one-dimensional drift scans at constant declination, whereas our deep slices were placed along the symmetry axis of each CHVC, 
making an interpretation in the sense of object symmetry more meaningful.

Another possible mechanism to produce the observed complex morphologies of most of our CHVCs is tidal interaction. Tidal 
forces, however, would act on both the compact cold cores and the diffuse warm envelopes in the same manner so that a misalignment 
between the core and the envelope, as in the case of head-tail CHVCs, cannot be explained by tidal interaction alone. But tidal 
forces could have shaped some of the irregular CHVCs in our sample. Clouds like \object{CHVC 221$-$88} or \object{CHVC 358$+$12} are highly elliptical, 
but they do not show any obvious misalignment between the diffuse envelope and the embedded cores, indicating that ram-pressure might 
not have been the principal process shaping these clouds.

Radiation pressure could also act on gas clouds and cause a misalignment between compact cores and diffuse envelopes. This 
process, however, can be excluded in the case of our CHVCs because it would require a strong radiation field as well as high density 
clumps with a significant amount of dust which has not yet been found in CHVCs. Wakker \& Boulanger (\cite{wakker3}) investigated 
parts of the HVC complexes A and M in IRAS observations, but they did not detect any dust in these HVCs, concluding that either the 
dust was too cold to be detected or that the dust abundance was below their detection limit. CHVCs have even lower \ion{H}{i} peak 
column densities than the large HVC complexes so that the presence of large amounts of dust is not expected if one assumes the same dust 
properties as for the low-velocity \ion{H}{i} clouds in the Galaxy. In addition, the strong radiation field should ionise the part of 
the CHVCs being exposed to the radiation. This should result in high H$\alpha$ intensities which are not consistent with the faint 
H$\alpha$ emission found by Tufte et al. (\cite{tufte}) and Putman et al. (\cite{putman2}).

\subsection{\label{sect_ram-pressure}\object{CHVC 017$-$25} as a prime example of ram-pressure interaction}

The most compelling example of a presumably interacting cloud, \object{CHVC 017$-$25}, is discussed in more detail at this point as 
the effects of ram-pressure distortion are quite pronounced and representative of other clouds in our sample. As already 
mentioned in Sect.~\ref{sect_chvc017-25}, the head-tail structure of \object{CHVC 017$-$25} does not seem very prominent at first glance. 
A more detailed analysis of the deep spectra along the apparent symmetry axis, however, provides evidence of a possible ram-pressure 
distortion of the entire cloud. Many spectra around the central part of \object{CHVC 017$-$25} show a more complex \ion{H}{i} line profile, 
indicating the superposition of two distinct gas components. Fig.~\ref{fig_chvc01_twogauss}~(a) presents an example spectrum in 
which a narrow line component ($\Delta v \approx 7 \; \mathrm{km \, s^{-1}}$ FWHM) of cold neutral gas ($T \lesssim 1100 \, 
\mathrm{K}$) seems to be superposed on a broad component ($\Delta v \approx 20 \; \mathrm{km \, s^{-1}}$ FWHM) of warm neutral gas 
($T \lesssim 9000 \, \mathrm{K}$). Therefore, we decomposed the \ion{H}{i} line profiles along the deep slice across \object{CHVC 017$-$25}, 
where possible, into two Gaussian components.

Of course, such a Gaussian decomposition is an approximation as we will never meet two completely distinct gas phases with 
uniform temperatures. As de Heij et al. (\cite{deheij}) have already pointed out, a decomposition of spectral lines into two 
Gaussian components, therefore, does not necessarily refer to two physically distinct systems. Nonetheless, the homogeneous 
physical parameters of the two line components justify the distinction of a cold and a warm gas phase as long as we only 
consider the qualitative properties of these two gas phases and their implications.

The results of the Gaussian decomposition of the spectral lines of \object{CHVC 017$-$25} are presented in Fig.~\ref{fig_chvc01_twogauss}~(b) 
-- (d). In Fig.~\ref{fig_chvc01_twogauss}~(b) we have plotted the line widths of the two gas components as a function of relative 
position along the deep slice. The two gas phases are clearly separated by their different line widths of $\Delta v \approx 6 \ldots 
10 \; \mathrm{km \, s^{-1}}$ FWHM for the cold gas and $\Delta v \approx 16 \ldots 26 \; \mathrm{km \, s^{-1}}$ FWHM in the case of the 
warm gas component. Fig.~\ref{fig_chvc01_twogauss}~(c) shows the column density profiles of both gas phases along the slice. The cold 
gas phase forms a compact head while the warm gas component is spatially much more extended so that we can speak of a compact, cold 
core embedded in a diffuse, warm envelope. The most remarkable result is that both gas phases are spatially separated from each other. 
At the south-eastern edge of the cloud only the narrow lines of the cold gas component can be seen. Towards the north-western edge we 
can only trace the warm gas phase which forms an extended, faint tail and justifies our classification of \object{CHVC 017$-$25} as a 
head-tail CHVC. This clear spatial separation of the cold core and the warm envelope could be explained if the 
envelope is being stripped off the compact core by the ram pressure of an ambient medium through which \object{CHVC 017$-$25} is moving.

In Fig.~\ref{fig_chvc01_twogauss}~(d) we have plotted the radial velocities of the two gas components across the deep slice. 
There is no clear gradient but a semi-periodic variation of radial velocities along the major axis of the cloud. At the four 
positions where both the cold and warm gas phases have been detected the velocity of the cold gas seems to be systematically 
higher (in absolute value) than that of the warm gas. This would be consistent with the idea of the cold core moving 
faster than the stripped-off warm envelope if we assume an infall of the cloud onto the Galaxy which is 
suggested by the highly negative radial velocity of $-171 \; \mathrm{km \, s}^{-1}$ in the GSR frame. But the velocity 
difference of about $1 \; \mathrm{km \, s^{-1}}$ between the core and the envelope is significantly smaller than the 
$2.6 \; \mathrm{km \, s^{-1}}$ velocity resolution of our spectra so that this result is not reliable.

\subsection{Distance of CHVCs}

The most controversial parameter of CHVCs is their distance. Distance is the key to most of the physical parameters 
like the \ion{H}{i} mass, the size, and, naturally, the overall distribution of the observed CHVC population within the Local Group.
As the distance of CHVCs cannot be determined directly we have to apply indirect distance estimates. Burton et al. (\cite{burton}) 
used an estimate of the thermal pressure at the interface of cold and warm neutral medium to obtain distances between 150 and 
850~kpc for a number of CHVCs observed with the Arecibo telescope. Such distances would place the CHVC population throughout 
the entire Local Group. Br\"uns et al. (\cite{bruens}) applied the virial theorem to a compact clump within \object{CHVC 125$+$41} to 
estimate a distance of the cloud of the order of 130~kpc. All these indirect distance determinations, however, require the 
assumptions of spherical symmetry and dynamical equilibrium of the cloud. We have seen that most of the 11 investigated CHVCs 
have a clearly non-spherical appearance and look highly disturbed so that in these cases simple distance estimates cannot be 
applied.

Evidence for a distribution of CHVCs within the Milky Way halo was found by Tufte et al. (\cite{tufte}). They detected 
H$\alpha$ emission towards four out of five CHVCs investigated with the Wisconsin H-Alpha Mapper. The H$\alpha$ intensities are 
comparable to those expected for clouds in the Galactic halo where the radiation field is strong enough to ionise the gas 
(see also Weiner et al. \cite{weiner}). The metagalactic radiation field throughout the Local Group would be too weak to account 
for the detected H$\alpha$ intensities. Putman et al. (\cite{putman2}) were able to confirm these results with the detection of 
H$\alpha$ emission towards two other CHVCs. These H$\alpha$ measurements suggest that CHVCs might constitute a circumgalactic 
population with distances of only some 10 kpc from the Galaxy.

\begin{figure}
\centering
  \includegraphics[width=8cm]{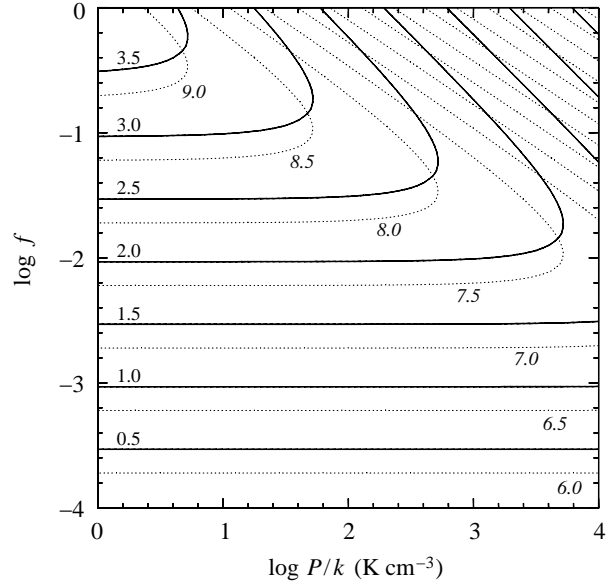}
  \caption{Constraining the distance of \object{CHVC 148$-$82}. Distance ($\log d / \mathrm{kpc}$, solid contours, labelled from $0.5$ to 
    $3.5$) and equilibrium mass ($\log M / M_{\odot}$, dotted contours, labelled from $6.0$ to $9.0$ but going down to $2.5$ in 
    the upper right corner of the diagram) plotted as a function of external pressure $P/k$ and \ion{H}{i} mass fraction $f \equiv 
    M_{\mathrm{\ion{H}{i}}} / M$ according to Eqn.~\ref{eqn_pressure2}. In the most reasonable parameter range of $\log f \approx 
    -2 \ldots 0$ and $\log P/k \approx 1 \ldots 3$ the distance of \object{CHVC 148$-$82} is not well constrained and has values between 
    30~kpc and 2~Mpc.}
  \label{fig_massenplot}
\end{figure}

Our only cloud with a spherically-symmetric appearance is \object{CHVC 148$-$82}. We can try to derive a rough distance estimate for 
this object by applying the following assumptions. Let us consider the cloud to be spherically-symmetric with radius $R$ and 
a constant mass density $\varrho = N_{\mathrm{\ion{H}{i}}} m_{\mathrm{H}} / (2 R f)$, where $N_{\mathrm{\ion{H}{i}}}$ is the 
central \ion{H}{i} column density, $m_{\mathrm{H}}$ is the mass of a hydrogen atom, and the factor $f \equiv M_{\mathrm{\ion{H}{i}}} 
/ M$ has been introduced to account for additional mass components such as ionised gas or dark matter which are not traced by 
the \ion{H}{i} column density. The total mass of the cloud is $M = (4/3) \pi R^3 \varrho$. Furthermore, we consider \object{CHVC 148$-$82} 
to be gravitationally bound and in dynamical equilibrium so that the virial theorem can be applied. Under the above assumptions the 
virial theorem reads
\begin{equation}
   M \langle v^2 \rangle = \frac{3}{5} \, \frac{G M^2}{R} \, , \label{eqn_virial_theorem}
\end{equation}
where $G$ is the gravitational constant, and $\langle v^2 \rangle$ denotes the mean-square velocity of the particles which is 
related to the observed FWHM $\Delta v$ of the spectral lines by $\langle v^2 \rangle = 3 \Delta v^2 / (8 \ln 2)$. We can now 
solve Eqn.~\ref{eqn_virial_theorem} for the distance $d$ of the cloud:
\begin{equation}
  d = \frac{15}{16 \pi \ln 2 \, G m_{\mathrm{H}}} \, \frac{f \Delta v^2}{N_{\mathrm{\ion{H}{i}}} \vartheta} \, . \label{eqn_distance}
\end{equation}
Here, $\vartheta$ denotes the angular radius of the cloud. In the case of \object{CHVC 148$-$82}, the average line width is $\Delta v \approx 
20 \; \mathrm{km \, s^{-1}}$ FWHM, the observed peak column density is $N_{\mathrm{\ion{H}{i}}} \approx 8 \cdot 10^{19} \; 
\mathrm{cm}^{-2}$, and the angular diameter amounts $2 \vartheta \approx 40'$. Inserting these values into Eqn.~\ref{eqn_distance} 
results in a distance estimate for \object{CHVC 148$-$82} of $d \approx f \cdot 10 \; \mathrm{Mpc}$. Considering the expected additional 
amount of helium of $M_{\mathrm{He}} \approx 0.4 M_{\mathrm{\ion{H}{i}}}$ (i.e. $f \approx 0.7$), \object{CHVC 148$-$82} would lie at a distance 
of about $7 \; \mathrm{Mpc}$ and would have a total mass of $3.2 \cdot 10^9 \; M_{\odot}$ and a radius of about $40 \; \mathrm{kpc}$. 
These parameters are highly implausible. A closer distance can be achieved by assuming a significant amount of additional mass 
(e.g. ionised gas or dark matter) or by rejecting the assumption of virialisation of \object{CHVC 148$-$82}. The problem that pure \ion{H}{i} 
CHVCs cannot be gravitationally stable at reasonable distances was discussed in detail by Sternberg et al. (\cite{sternberg}). They 
consider the possibility of an additional pressure support of CHVCs by the presence of a surrounding medium. The presence of such an 
ambient medium in which CHVCs are embedded is illustrated by the majority of disturbed clouds among our sample. We can 
account for an external pressure support by inserting an additional pressure term into the virial theorem which then reads 
\begin{equation}
  M \langle v^2 \rangle = \frac{3}{5} \, \frac{G M^2}{R} + 3 P V \, . \label{eqn_pressure}
\end{equation}
Here, $V = (4/3) \pi R^3$ is the volume of the cloud, and $P$ denotes the external pressure. This equation, however, is only 
valid for clouds without any dark matter that is not governed by an external pressure component. Solving Eqn.~\ref{eqn_pressure} 
for $P$ yields 
\begin{equation}
  P = \frac{m_{\mathrm{H}} N_{\mathrm{\ion{H}{i}}} \Delta v^2}{16 \ln 2 \, f \vartheta d} - \frac{\pi G m_{\mathrm{H}}^2 
  N_{\mathrm{\ion{H}{i}}}^2}{15 f^2} \, . \label{eqn_pressure2} 
\end{equation}
According to Eqn.~\ref{eqn_pressure2}, the distance of the cloud is influenced by two free parameters, the external pressure 
$P$ and the \ion{H}{i} mass fraction $f$. In Fig.~\ref{fig_massenplot} we have plotted the mass $M$ (dotted contours) and the 
distance $d$ (solid contours) of \object{CHVC 148$-$82} as a function of these two parameters according to Eqn.~\ref{eqn_pressure2}. For 
very small $f \lesssim 10^{-2}$ both distance and mass of the cloud become independent of the external pressure because 
stabilisation of the cloud is by far dominated by the mass. In this case, the distance would be $d \lesssim 100 \; \mathrm{kpc}$ 
and the mass $M \lesssim 5 \cdot 10^7 \; M_{\odot}$. For more realistic values of $f$ distance and mass strongly depend on the 
strength of the external pressure. In the most reasonable parameter range of $\log f \approx -2 \ldots 0$ and $\log P/k \approx 
1 \ldots 3$ the distance of \object{CHVC 148$-$82} is not well constrained and has values between 30~kpc and 2~Mpc. These results show that 
we cannot obtain any reliable distance estimate for \object{CHVC 148$-$82} due to the major uncertainties in parameter space in combination 
with a number of doubtful assumptions such as constant mass density or virialisation of the cloud.

In other words, within a realistic range of values for the external pressure and the \ion{H}{i} mass fraction we can derive any 
distance within the Local Group for \object{CHVC 148$-$82}. This makes earlier distance estimates, placing CHVCs throughout the entire Local 
Group, doubtful as they are governed by similar assumptions and uncertainties. On the other hand, we observe interaction 
effects in our data. Head-tail structures and bow-shock shapes are suggestive of ram-pressure interaction between CHVCs 
and the surrounding medium which may indicate a distribution of CHVCs in the vicinity of the Milky Way. Sternberg et al. (\cite{sternberg}) 
simulated a circumgalactic population of CHVCs with distances of $150 \; \mathrm{kpc}$ from the Milky Way. If we consider the same 
distance for \object{CHVC 148$-$82} we would need an external pressure of $P / k \approx 180 \; \mathrm{K \, cm^{-3}}$ according to 
Eqn.~\ref{eqn_pressure2} to stabilise the cloud in the case of no additional dark matter and gas content except helium. This seems 
to be a rather high external pressure for such large distances from the Galaxy. Any dark matter content, however, will additionally 
stabilise \object{CHVC 148$-$82} and decrease the required amount of external pressure. 

Furthermore, there are indications for the existence of a hot ionised intergalactic medium which could provide an additional 
pressure stabilisation of CHVCs. Rasmussen et al. (\cite{rasmussen}) used X-ray absorption lines of \ion{O}{vii}, \ion{O}{viii}, 
\ion{C}{vi}, and \ion{Ne}{ix} detected against three active galactic nuclei to constrain the physical properties of the absorbing 
gas. They obtained an upper limit for the electron density of the ionised medium of $n_{\mathrm{e}} < 2 \cdot 10^{-4} \; 
\mathrm{cm}^{-3}$, corresponding to a thermal pressure of $P / k \lesssim 770 \; \mathrm{K \, cm^{-3}}$ with a gas temperature of 
$2 \cdot 10^6 \; \mathrm{K}$. Following limitations in the possible scale lengths of the absorber, Rasmussen et al. 
(\cite{rasmussen}) were able to locate the absorbing gas within the Local Group with a lower limit for the scale length of about 
140~kpc. Even if the pressure provided by this ionised intergalactic medium is not sufficient to support CHVCs alone, a 
combination of external pressure and dark matter content could stabilise clouds like \object{CHVC 148$-$82} even at distances of the order 
of $100 \; \mathrm{kpc}$ from the Galaxy.

The above results and considerations lead to a consistent picture in which CHVCs might constitute a circumgalactic population 
with typical distances of the order of 100~kpc from the Milky Way. At such distances, CHVCs could be embedded in a hot ionised 
intergalactic gas or an extended Galactic corona and be distorted by ram-pressure interaction with this ambient medium. Furthermore, 
the external pressure provided by the ambient medium could support CHVCs in addition to their own gravitational potential.

\section{\label{sect_summary}Summary}

We have mapped 11 compact high-velocity clouds (CHVCs) in the 21-cm line emission of neutral, atomic hydrogen, using the Effelsberg 
100-m radio telescope. We obtained a baseline rms of roughly $\sigma_{\mathrm{rms}} = 50 \; \mathrm{mK}$ at the original velocity 
resolution of $2.6 \; \mathrm{km \, s^{-1}}$, leading to a 1-sigma \ion{H}{i} column density detection limit of 
about $2.4 \cdot 10^{17} \; \mathrm{cm}^{-2}$ per spectral channel. These maps allow us to examine 
the overall morphology of the clouds. In addition, we have obtained deep spectra along the symmetry axis across each CHVC with 
denser angular sampling. Along these deep slices we reach a baseline rms of $\sigma_{\mathrm{rms}} \approx 25 \ldots 35 \; \mathrm{mK}$ 
at the original velocity resolution of $2.6 \; \mathrm{km \, s^{-1}}$, resulting in an average 1-sigma \ion{H}{i} column density 
detection limit of about $1.2 \ldots 1.7 \cdot 10^{17} \; \mathrm{cm}^{-2}$ per spectral channel. These deep slices allow us to extract 
the column density profile in great detail as well as the velocity and line width gradient across each cloud.

The most outstanding result of our observations is the complexity of the investigated CHVCs. The numerous head-tail structures 
and bow-shock shapes point out that the head-tail structure of \object{CHVC 125$+$41} found by Br\"uns et al. (\cite{bruens}) is not unique. 
Among our 11 objects we found only one cloud with a spherically-symmetric appearance. Four CHVCs have a head-tail structure, two 
CHVCs show a bow-shock shape, and the remaining four clouds are irregular. The complex morphologies of most of our 11 CHVCs 
suggest that they may be disturbed by the ram pressure of the surrounding medium. In cases like \object{CHVC 017$-$25}, the diffuse 
envelope of warm neutral gas appears to be stripped off the compact cold core of the cloud, forming a more or less extended, faint 
\ion{H}{i} tail with column densities of typically $N_{\mathrm{\ion{H}{i}}} \lesssim 10^{19} \; \mathrm{cm}^{-2}$. Some of the 
irregular CHVCs in our sample are highly elliptical but do not show a noticeable misalignment between the diffuse envelope and the 
embedded clumps. In these cases, tidal interactions instead of ram pressure might have played a significant role in shaping the clouds.

The existence of an ambient medium around CHVCs can also solve the problem of stability. If CHVCs were only made of \ion{H}{i} 
they would not be gravitationally stable at distances of less than a few Mpc. At such large distances, however, CHVCs would be 
extremely large and diffuse objects with \ion{H}{i} masses of a few times $10^9 \; M_{\odot}$. Additional mass components like 
ionised gas, molecular gas or dark matter could solve this distance problem and help to stabilise CHVCs. Another option, however, 
is stabilisation by external pressure. We showed that at a distance of $150 \; \mathrm{kpc}$ an external pressure of $P / k 
\approx 180 \; \mathrm{K \, cm^{-3}}$ would be sufficient to stabilise objects like \object{CHVC 148$-$82} even without any additional 
mass components except for the expected associated helium content.

The above considerations show that our observations are consistent with a circumgalactic population of CHVCs with distances of 
the order of $100 \; \mathrm{kpc}$. At such distances, CHVCs can be embedded in the gaseous environment of an extended Galactic 
corona or an ionised intergalactic medium which not only stabilises the clouds but which can also provide the necessary ram 
pressure to account for the observed complex morphologies of most of our CHVCs.

\begin{acknowledgements}
  We wish to thank our anonymous referee for many useful comments which helped to improve this paper. 
  \mbox{T. W.} acknowledges support by the Deutsche Forschungsgemeinschaft (German Research Foundation) through project number 
  \mbox{KE757/4--1}. The results presented in this paper are based on observations with the 100-m telescope of the MPIfR 
  (Max-Planck-Institut f\"ur Radioastronomie) at Effelsberg.
\end{acknowledgements}


\begin{thebibliography}{}
\bibitem[1995]{banks}
  Banks, T., Dodd, R. J., \& Sullivan, D. J. 1995, MNRAS, 272, 821
\bibitem[2001]{barnes}
  Barnes, D. G., Staveley-Smith, L., de Blok, W. J. G., et al. 2001, MNRAS, 322, 468
\bibitem[1999]{blitz}
  Blitz, L., Spergel, D. N., Teuben, P. J., Hartman, D., \& Burton, W. B. 1999, \apj, 514, 818
\bibitem[1999]{braun}
  Braun, R. \& Burton, W. B. 1999, \aap, 341, 437
\bibitem[2000]{braun2}
  Braun, R. \& Burton, W. B. 2000, \aap, 354, 853
\bibitem[2000]{bruens2}
  Br\"uns, C., Kerp, J., Kalberla, P. M. W., \& Mebold, U. 2000, \aap, 357, 120
\bibitem[2001]{bruens}
  Br\"uns, C., Kerp, J., \& Pagels, A. 2001, \aap, 370, L26
\bibitem[2004]{bruens3}
  Br\"uns, C., Kerp, J., Staveley-Smith, L., et al. 2004, \aap, in press
\bibitem[2001]{burton}
  Burton, W. B., Braun, R., \& Chengalur, J. N. 2001, \aap, 369, 616
\bibitem[1997]{hartmann}
  Hartmann, D. \& Burton W. B. 1997, Atlas of Galactic Neutral Hydrogen, Cambridge University Press
\bibitem[2002a]{deheij}
  de Heij, V., Braun, R., \& Burton W. B. 2002a, \aap, 391, 67
\bibitem[2002b]{deheij2}
  de Heij, V., Braun, R., \& Burton W. B. 2002b, \aap, 391, 159
\bibitem[2002c]{deheij3}
  de Heij, V., Braun, R., \& Burton W. B. 2002c, \aap, 392, 417
\bibitem[1980]{kalberla}
  Kalberla, P. M. W., Mebold, U., \& Reich, W. 1980, \aap, 82, 275
\bibitem[1982]{kalberla2}
  Kalberla, P. M. W., Mebold, U., \& Reif, K. 1982, \aap, 106, 190
\bibitem[1999]{klypin}
  Klypin, A., Kravtsov, A. V., Valenzuela, O., \& Prada, F. 1999, \apj, 522, 82
\bibitem[1999]{moore}
  Moore, B., Ghigna, S., Governato, G., Lake, G., Quinn, T., et al. 1999, \apj, 524, L19
\bibitem[1963]{muller}
  Muller, C. A., Oort, J. H., \& Raimond, E. 1963, C. R. Acad. Sci. Paris, 257, 1661
\bibitem[2004]{pisano}
  Pisano, D. J., Barnes, D. G., Gibson, B. K., Staveley-Smith, L., Freeman, K. C., \& Kilborn, V. A. 2004, \apj, 610, L17
\bibitem[2003]{putman2}
  Putman, M. E., Bland-Hawthorn, J., Veilleux, S., et al. 2003, ApJ, 597, 948
\bibitem[2002]{putman}
  Putman, M. E., de Heij, V., Staveley-Smith, L., et al. 2002, \aj, 123, 873
\bibitem[2001]{quilis}
  Quilis, V. \& Moore, B. 2001, \apj, 555, L95
\bibitem[2003]{rasmussen}
  Rasmussen, A., Kahn, S. M., \& Paerels, F. 2003, in ASSL Conf. Proc. Vol. 281, The IGM/Galaxy Connection: The Distribution of 
  Baryons at $z = 0$, eds. J. L. Rosenberg \& M. E. Putman, 109
\bibitem[2002]{sternberg}
  Sternberg, A., McKee, C. F., \& Wolfire, M. G. 2002, \apjs, 143, 419
\bibitem[2002]{tufte}
  Tufte, S. L., Wilson, J. D., Madsen, G. J., Haffner, L. M., \& Reynolds, R. J. 2002, \apj, 572, L153
\bibitem[1986]{wakker3}
  Wakker, B. P. \& Boulanger, F. 1986, \aap, 170, 84
\bibitem[1991]{wakker}
  Wakker, B. P. 1991, \aap, 250, 499
\bibitem[1997]{wakker2}
  Wakker, B. P. \& van Woerden, H. 1997, \araa, 35, 217
\bibitem[2001]{weiner}
  Weiner, B. J., Vogel, S. N., \& Williams, T. B. 2001, in ASP Conf. Proc. Vol. 240, Gas and Galaxy Evolution, eds. J. E. Hibbard, 
  M. Rupen, \& J. H. van Gorkom, 515
\bibitem[2003]{westmeier}
  Westmeier, T. 2003, Diploma thesis, Rheinische Friedrich-Wilhelms-Universit\"at Bonn
\bibitem[2000]{zwaan}
  Zwaan, M. 2000, Ph. D. thesis, Rijksuniversiteit Groningen
\end{thebibliography}
\end{document}